\documentclass[9pt,twocolumn,twoside]{pnas-new-preprint}
\usepackage{styleFile}
\templatetype{pnasresearcharticle} 

\title{A single parameter can predict surfactant impairment of superhydrophobic drag reduction}

\author[a]{Fernando Temprano-Coleto}
\author[a]{Scott M. Smith} 
\author[b]{Fran{\c{c}}ois J. Peaudecerf}
\author[c]{Julien R. Landel}
\author[a]{Fr{\'e}d{\'e}ric Gibou}
\author[a,1]{Paolo Luzzatto-Fegiz}

\affil[a]{Department of Mechanical Engineering, University of California, Santa Barbara, CA 93106, USA}
\affil[b]{Institute of Environmental Engineering, Department of Civil, Environmental and Geomatic Engineering, Eidgenössische Technische Hochschule (ETH) Zürich, 8093 Z{\"{u}}rich, Switzerland}
\affil[c]{Department of Mathematics, Alan Turing Building, University of Manchester, Oxford Road, Manchester M13~9PL, United Kingdom}

\leadauthor{Temprano-Coleto} 

\significancestatement{Trace surfactants, unavoidable in applications, can impair the drag reduction achieved by superhydrophobic surfaces (SHS), as Marangoni stresses immobilize the air-water interface. It is not known how SHS impairment depends on surfactant type and concentration, flow velocity, and SHS geometry; as a result, mitigation strategies are still needed. We introduce a model of this phenomenon, and perform simulations and experiments.  We find that the interface can be mobilized if it is longer than a critical length scale, which is determined by the surfactant properties, essentially independently of flow velocity. SHS impairment is thereby predicted from a single parameter, namely the ratio of interface length and mobilization scale, providing fundamental insight and practical guidance to achieve superhydrophobic drag reduction.
}

\authorcontributions{Author contributions: 
F.T.-C., F.J.P., J.R.L., F.G., and P.L.-F. designed research; 
F.T.-C., S.M.S., F.J.P., J.R.L., and P.L.-F. performed research;
F.T.-C., S.M.S., F.J.P., J.R.L., and P.L.-F. analyzed data;
and F.T.-C., F.J.P., J.R.L., F.G., and P.L.-F. wrote the paper.
}
\correspondingauthor{\textsuperscript{1}To whom correspondence should be addressed. E-mail: pfegiz@ucsb.edu}


\begin{abstract}
Recent experimental and computational investigations have shown that trace amounts of surfactants, unavoidable in practice, can critically impair the drag reduction of superhydrophobic surfaces (SHSs), by inducing Marangoni stresses at the air-liquid interface. However, predictive models for realistic SHS geometries do not yet exist, which has limited the understanding and mitigation of these adverse surfactant effects. To address this issue, we derive a model for laminar, three-dimensional flow over SHS gratings, as a function of geometry and soluble surfactant properties, which together encompass ten dimensionless groups. We establish that the grating length $g$ is the key geometric parameter, and predict that the ratio between actual and surfactant-free slip increases with $g^2$. Guided by our model, we perform synergistic numerical simulations and microfluidic experiments, finding good agreement with the theory as we vary surfactant type and SHS geometry. Our model also enables the estimation, based on velocity measurements, of \textit{a priori} unknown properties of surfactants inherently present in microfluidic systems. For SHSs, we show that surfactant effects can be predicted by a single parameter, representing the ratio between the grating length and the interface length scale beyond which the flow mobilizes the air-water interface. This mobilization length is more sensitive to the surfactant chemistry than to its concentration, such that even trace-level contaminants may significantly increase drag if they are highly surface-active. These findings advance the fundamental understanding of realistic interfacial flows, and provide practical strategies to maximize superhydrophobic drag reduction.
\end{abstract}

\begin{document}

\maketitle
\thispagestyle{firststyle}
\ifthenelse{\boolean{shortarticle}}{\ifthenelse{\boolean{singlecolumn}}{\abscontentformatted}{\abscontent}}{}

\dropcap{S}uperhydrophobic surfaces (SHSs) have the potential to yield enormous technological benefits in fields ranging from microfluidics to maritime transportation, primarily due to their ability to reduce drag \cite{Rothstein:2010im}. Through a combination of hydrophobic chemistry and microscopic surface patterning, these substrates are able to retain a superficial layer of air, thereby producing an apparent slip when in contact with a liquid flow \cite{Lauga07}. Early theoretical work \cite{Philip_ZAMP_1972a,Philip_ZAMP_1972b,Lauga_Stone_JFM_2003} modeled the air pockets trapped within these textures as flat boundaries with no shear, predicting large drag reductions in the laminar regime. Although early experiments found promising levels of drag reduction \cite{Ou2004-vk,Ou2005-ph,Truesdell2006-nj,Lee2008-mg}, subsequent studies measured a reduced or even non-existent slip \cite{Kim2012-iw,Bolognesi2014-vw,Schaffel2016-mh}, pointing at the interfacial stresses induced by surface-active contaminants as one possible cause of this discrepancy. 
Recently, independent experimental studies have reported time-dependent and spatially complex interfacial dynamics that unequivocally demonstrate the importance of surfactant-induced stresses on SHSs \cite{peaudecerf17,Song2018-uw}. Theoretical and computational works have confirmed the extent to which trace amounts of these surface-active contaminants can reduce slip \cite{landel20,Li20,Baier21,Sundin22}. This slip reduction is also consistent with broader findings for small-scale multiphase flows, where environmental levels of surfactants, often extremely difficult to avoid or control, play a central role \cite{Manikantan20}\revised{; prominent examples are given by small bubbles rising in water (e.g.~\cite{bond28,frumkin47,levich62,Palaparthi2006-di} and references therein) or bubbles probed by atomic force microscopy \cite{Manor2008,Maali2017}. These flows have been understood through models that include surfactants, sometimes at trace levels that are undetectable by traditional surface tension measurements.
}

For SHS textures, the concentration gradients that induce Marangoni stresses appear in the streamwise direction, owing to stagnation points at the downstream ends of the interfaces, where advected surfactants accumulate (see \figref{fig:fig1} A and B). Modeling this physical mechanism for realistic SHS geometries is challenging. In addition to the four coupled partial differential equations governing the physics and the ten associated dimensionless numbers (detailed below), there is a major difficulty stemming from the alternating slip/no-slip boundary conditions at the edges where the fluid interface meets the solid substrate. The resulting spatially complex flows constitute a challenge for analytical progress. 
For this reason, models with surfactants only considered two-dimensional flows over \emph{transverse} SHS gratings, as this is the simplest geometry that captures detrimental surfactant effects \cite{landel20,Baier21}. A more realistic configuration is that of \emph{streamwise} gratings, which are widely used \cite{Ou2004-vk,tsai09,Bolognesi2014-vw,peaudecerf17}, owing to their potential for very high drag reduction in surfactant-free conditions. Gratings have been modeled as infinitely long in surfactant-free theories \citep{Philip_ZAMP_1972a,Lauga_Stone_JFM_2003,teo09,feuillebois09,Asmolov2012}; 
however, modeling surfactant effects requires considering \emph{finite} streamwise gratings with stagnation points, leading to a three-dimensional (3D) flow. Theories of realistic gratings inclusive of surfactant are still needed.

Here, we introduce a theory for 3D flow over streamwise SHS gratings with surfactants, by coupling a new hydrodynamic solution (for slender, finite gratings with arbitrary shear at the interface) with a scaling analysis of surfactant dynamics (for soluble, dilute surfactants). We use our model to design experiments and simulations where the slip velocity varies across three orders of magnitude, relative to surfactant-free conditions, and thereby achieve a direct comparison between theory and experiments for realistic SHSs. Our theory can also use velocity measurements to estimate physicochemical parameters of unknown, trace-level surfactants, which are inevitable both in natural and artificial settings. Although the general problem comprises ten dimensionless groups, we show that impairment by trace surfactant is approximately controlled by a single parameter, which depends on surfactant type and concentration, and is independent of flow velocity. Since surface-active molecules are naturally released by polymers widely used in microfabrication \cite{Regehr09,Hourlier-Fargette17,Hourlier-Fargette2018-tu,carter:20}, we expect these results to be valuable over a broad range of fundamental and applied microfluidic research.

\section*{SHS model for 3D flow with surfactants}

\begin{figure}[t!]
\vspace{-10pt}
  \centering
  \captionsetup{farskip=0pt, nearskip=0pt}
  \includegraphics[width=\linewidth]{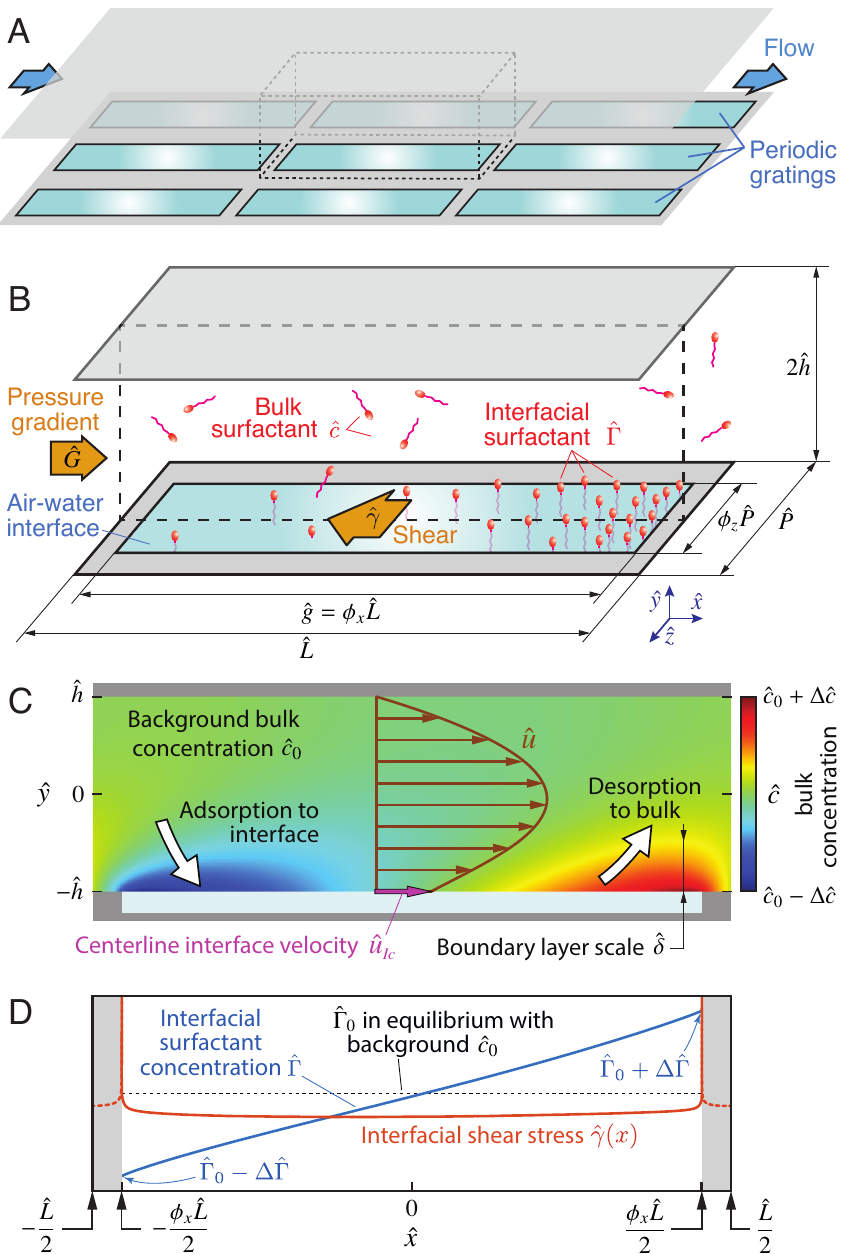}
  \caption{\label{fig:sketch}\revised{(A) Diagram showing the plane channel flow arrangement studied, with an array of slender periodic streamwise gratings on the bottom  wall. (B) Unit cell of the SHS (also shown with dashed lines in panel A), periodic in $\hx$ and $\hz$, illustrating the downstream accumulation of surfactant. (C) Streamwise cross section at mid-grating ($\hz=0$, also shown with dashed lines in panel B), showing field of bulk surfactant from a representative simulation, with adsorption/desorption regions at the upstream/downstream ends, respectively. (D) Interface concentration (blue) and shear stress (red), for the same $\hz=0$ cross section. Throughout the article, hats denote dimensional quantities.}
  }\label{fig:fig1}
\end{figure}

We consider steady, laminar flow driven by a mean pressure gradient $\hG$ across a channel of half-height $\hh$, where hats denote dimensional quantities. The bottom of the channel is lined with a periodic pattern of slender, rectangular gratings \revised{(Fig.~\ref{fig:fig1}A)}. Each gas-liquid interface (the `{plastron}') is assumed flat. 
Due to the periodicity of the array in the streamwise and spanwise directions, we focus on a unit cell consisting of one grating and its surrounding ridges, as depicted in \revised{Fig.~\ref{fig:fig1}A}. %
The streamwise, wall-normal, and spanwise directions are $\hx$, $\hy$ and $\hz$, respectively, with the coordinate origin at the center of the unit cell \revised{(Fig.~\ref{fig:fig1}B)}. 

We leverage the disparity of scales between the length $\hL$ and the half-height $\hh$ (see \revised{Fig.~\ref{fig:fig1}B}), and define a small parameter $\eps=\hh/\hL\ll{1}$, which is dimensionless and thus written without hats. Differently from the classic Hele-Shaw flow approximation \cite{feuillebois09}, here we do not assume that the spanwise length scale (the pitch $\hP$) is much larger than $\hh$, since, in microfluidic applications $\hh$ and $\hP$ are of the order of tens of micrometers, whereas $\hL$ ranges in the millimeter or centimeter scale \cite{Ou2004-vk,Ou2005-ph,tsai09,Bolognesi2014-vw}. Consequently, we define the nondimensional coordinates $x=\hx/\hL$, $y=\hy/\hh=\hy/(\eps\hL)$ and $z=\hz/(\eps\hL)$. Incompressibility implies that the flow is approximately unidirectional, with the dominant streamwise velocity scaling as $\hu\sim\hU$, whereas the wall-normal and spanwise components scale as $\hv\sim\eps\hU$ and $\hw\sim\eps\hU$. The velocity scale is $\hU=\hh^2\hG/\visc$, with $\visc$ the dynamic viscosity. At leading order in $\eps$, the Stokes equations for the flow are $\partial_{yy}u + \partial_{zz}u = \partial_{x}p$ and $\partial_{y}p=\partial_{z}p=0$ (see \SM\textit{ Flow field derivation}), where $u(x,y,z)=\hu/\hU$ and $p(x)=\hp/(\hG\hL)$ are the dimensionless streamwise velocity and pressure. The unidirectional nature of this leading-order flow is a good approximation far from the downstream and upstream edges of the plastron, i.e.~where $|x\pm\gfx/2|\gg\eps$, with $\gfx$ the streamwise gas fraction (see \revised{\figref{fig:fig1}B}). Therefore, the asymptotic expansion in $\eps$ is {singular}, as is common for thin-gap approximations \cite{Leal07}. Since we consider slender gratings with $\eps\ll{1}$, the regions of validity represent most of the domain and useful approximations of both local and integrated quantities can be obtained.

No-slip boundary conditions $u=0$ are imposed at solid walls and ridges. The interface imposes a Marangoni stress $\hat{\gamma}$ determined by the local gradient of interfacial (adsorbed) surfactant.
This stress is independent of transverse direction $z$ at leading order in $\eps$ (see \SM\textit{ Flow field derivation}), and thus the plastron boundary condition is $\left.\partial_y{u}\right|_I=\gMa$, where $\gMa=\hat{\gamma}/(\visc\hU/\eps\hL)$ and the subscript $I$ denotes conditions at the interface. 

\begin{figure*}[b!]
   \centering
   \captionsetup{farskip=0pt, nearskip=0pt}
   \includegraphics[]{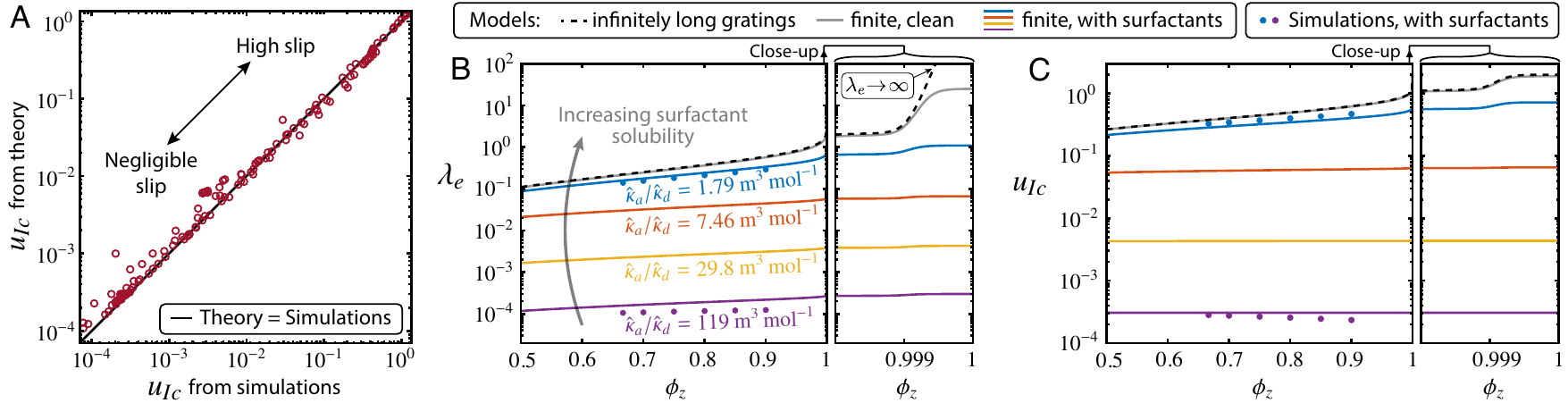}
   \caption{\revised{(A) Comparison of the centerline slip velocity from 155 numerical simulations with  our model prediction \equaref{eq:u_Ic}. (B) Effective slip length and (C) centerline slip velocity as a function of the spanwise gas fraction $\gfz$, for a fixed surfactant concentration $\hcz$ and varying surfactant solubility ($\hka/\hkd$). Unless noted, the parameters are as in \textit{Table~SI}}. 
   \label{fig:fit_agreement}\label{fig:lambda_vs_GF}\label{fig:uIc_vs_GF}
   }
   \label{fig:theory_sim}
\end{figure*}

Note that, for finite gratings, the pressure gradient is not constant in $x$, and $p(x)$ must be determined from two integral constraints. First, the volume flow rate $Q = \int_{-P/2}^{P/2}\int_{-1}^{1}u(x,y,z)\,\mathrm{d}y\,\mathrm{d}z$ \revised{(where $P=\hP/\hh$ is the normalized pitch, see \figref{fig:fig1}B)} must be independent of $x$, to satisfy mass conservation. Second, the pressure drop across the whole unit cell must match the imposed mean pressure gradient, such that $\int_{-1/2}^{1/2}\partial_x{p}(x)\,\mathrm{d}x = -1$. These two conditions lead to an expression for the flow field (as detailed in \SM\textit{ Flow field derivation}),
\begin{equation} \label{eq:flow_field}
\revised{
u(x,y,z) = 
  \begin{cases}
       \dfrac{\left[1+\qdinf(1-\gfx)\gMaavg\right]u_P(y)+\left[1-\gMaavg\right]u_d^{\infty}(y,z)}{1+\qdinf(1-\gfx)} \\[10pt]
       \hspace{8pt}-\dfrac{\gMa-\gMaavg}{1+\qdinf}\left[\qdinf u_P(y) - u_d^{\infty}(y,z) \right]
       ,\;\text{  if }|x| < \dfrac{\gfx}{2} \\[18pt]
      \left[ \dfrac{1+\qdinf(1-\gfx\gMaavg)}{1+\qdinf(1-\gfx)} \right] u_P(y)   ,\;\text{  if }\dfrac{\gfx}{2} < |x| \leq \dfrac{1}{2},
  \end{cases}
  }
\end{equation}
where $u_P(y)=(1-y^2)/2$ is the Poiseuille profile and $u_d^\infty(y,z)$ is the deviation from $u_P(y)$ in the \emph{infinite-grating case}, where the interface has no stagnation points and thus surfactant effects are absent. In other words, if $\gfx=1$ then $u(y,z)=u_P(y)+u_d^\infty(y,z)$ and $\gMa=0$, where $u_d^\infty(y,z)$ is known from previous studies \cite{Philip_ZAMP_1972a,Teo2008-pe}. In \equaref{eq:flow_field}, $\qdinf=3\Qdinf/(2P)$, where $\Qdinf = \int_{-P/2}^{P/2}\int_{-1}^{1}u_d^{\infty}(y,z)\,\mathrm{d}y\,\mathrm{d}z$, and $\gMaavg = \frac{1}{\gfx} \int_{-\gfx/2}^{\gfx/2}\gMa\,\mathrm{d}x$ is the average Marangoni shear across the plastron, which varies between 0, for a clean interface, and 1 for a fully immobilized interface. Equation~\eqref{eq:flow_field} provides the velocity field as a linear combination of two known simpler solutions ($u_P$ and $u_d^\infty$), relying on parameters which are either prescribed ($\gMa$) or known from the infinite-grating problem ($q_d^\infty$).

The flow field is linked to the surfactant dynamics via $\gamma$, which is found from the equations for soluble surfactant (see \SM\textit{ Governing equations})
\begin{subequations}
  \label{eq:surfactant}
  \begin{align} 
     & u\pd{c}{x}{} + v\pd{c}{y}{} + w\pd{c}{z}{} = \dfrac{1}{\eps{Pe}} \left( \eps^2\pd{c}{x}{2} + \pd{c}{y}{2} + \pd{c}{z}{2} \right),        \label{eq:surfactant:bulk}\\
     & \pd{(u\Gam)}{x}{} + \pd{(w\Gam)}{z}{}     = \dfrac{1}{\eps{Pe_I}} \left( \eps^2\pd{\Gam}{x}{2} +\pd{\Gam}{z}{2}\right) + \dfrac{Bi}{\eps}\left(c_I-\Gam \right), \label{eq:surfactant:interface}\\
     & \left.\pd{c}{y}{}\right|_I         = Da\left(c_I-\Gam\right)\text{\hspace{26pt}  at the interface},                            \label{eq:surfactant:kinetics}\\
     & \left.\pd{u}{y}{}\right|_I = \gMa     = \eps k Ma \pd{\Gam}{x}{} \text{\quad  at the interface}.                           \label{eq:surfactant:marangoni}
  \end{align}
\end{subequations}

\revised{Equations \eqref{eq:surfactant:bulk} and \eqref{eq:surfactant:interface} describe the transport of the nondimensional \emph{bulk} and \emph{interfacial} surfactant concentrations $c=\hat{c}/\hcz$ and $\Gam=\hGam/\hGamz$, respectively, where $\hcz$ is the background bulk concentration and $\hGamz$ is the equilibrium interfacial concentration (Fig.~\ref{fig:fig1}C,D).} The adsorption and desorption kinetics are modeled through \eqref{eq:surfactant:kinetics} and the last term in \eqref{eq:surfactant:interface}, whereas the Marangoni boundary condition \eqref{eq:surfactant:marangoni} relates the shear stress to the gradient of surfactant concentration at the interface. 
Six dimensionless groups control the surfactant dynamics in \eqsref{eq:surfactant}. The bulk and interface P{\'e}clet numbers are $Pe=\hh\hU/\hD$ and $Pe_I=\hh\hU/\hDI$, where $\hD$ and $\hDI$ are the bulk and interface diffusivities. The Marangoni number $Ma=\ns\hR\hT\hGamm/(\visc\hU)$ depends on the maximum interfacial packing concentration $\hGamm$, the ideal gas constant $\hR$, the temperature $\hT$ and a parameter $\ns$ quantifying the effects of salinity. The Biot $Bi=\hh\hkd/\hU$ and Damk{\"o}hler $Da=\hh\hka\hGamm/\hD$ numbers parameterize the effect of kinetics, where $\hka$ and $\hkd$ the adsorption and desorption rate constants. The normalized \revised{background concentration is $k=\hka\hcz/\hkd$, and can be related to the interfacial concentration at kinetic equilibrium through $k=\hGamz/\hGamm$.}  
These six dimensionless groups, in addition to four geometrical parameters $\gfx$, $\gfz$, $P$ and $g=\hg/\hh=\gfx/\eps$, fully describe the flow and surfactant transport problem.

A scaling analysis of \eqsref{eq:surfactant}, similar to the one performed in \cite{landel20} for transverse gratings, leads to an expression for $\gMaavg$. 
The derivation, which can be found in \SM\textit{ Scaling theory for surfactant transport}, is based on the assumption of low normalized concentration ($k\ll{1}$), which justifies the choice of Henry kinetics \cite{Manikantan20} in Eq.~\eqref{eq:surfactant}, \revised{and is the case in applications unless substantial amounts of surfactant are deliberately added \cite{Schaffel2016-mh,peaudecerf17,Song2018-uw}. The stress is approximated as spatially uniform, i.e.~$\gMa\approx\gMaavg$, as has been found to be the case in small-scale applications \cite{landel20}, such that the term on the second line of \eqref{eq:flow_field} is negligible.}

To quantify slip and enable comparison with experiments, we use the \emph{centerline slip velocity} $\uIc$, \revised{defined as the velocity along the centerline of the interface,  $\uIc=u(x,y=-1,z=0)$, as illustrated in Fig.~\ref{fig:fig1}C. We use $\uIc$ because} it can be measured with greater ease and accuracy than the \revised{local} slip length $\lambda=u_I/\left.\partial_y{u}\right|_I$, which requires estimation of velocity gradients \cite{Schaffel2016-mh}. From the combination of the flow field from \eqqref{eq:flow_field} and the expression for $\gMaavg$ obtained from the scaling of \eqsref{eq:surfactant}, we obtain
\begin{align}
  \dfrac{\uIc}{\uIcc} &= 1 - \dfrac{ a_1\,{k}\,{Ma}\,\uIcc }{ a_1\,{k}\,{Ma}\,\uIcc + a_2\dfrac{Bi\,g^2}{1+\bl{Da}} + \dfrac{1}{Pe_I}}.
  \label{eq:u_Ic}
  \end{align}
 \revised{Our model also yields an expression for the  \textit{effective slip length} $\lame$, defined through the wall-averaged Navier slip boundary condition that would result in the same flow rate as alternating no-slip/slip boundary conditions on the bottom wall \cite{Lauga_Stone_JFM_2003, Teo2008-pe,Rothstein:2010im,landel20}, }
  \begin{align}
  \revised{\dfrac{\lambda_e}{\lambda_e^\text{clean}}} &= \revised{1 - \dfrac{ a_1\,{k}\,{Ma}\,\uIcc \left(1+\dfrac{\lambda_e^\text{clean}}{2}\right)  }{ a_1\,{k}\,{Ma}\,\uIcc \left(1+\dfrac{\lambda_e^\text{clean}}{2}\right) + a_2\dfrac{Bi\,g^2}{1+\bl{Da}} + \dfrac{1}{Pe_I}}.}
  \label{eq:lambdaE}
\end{align}
\revised{In Eqs.~\eqref{eq:u_Ic} and \eqref{eq:lambdaE},} $\bl=\hat{\delta}/\hh$ is the concentration boundary layer thickness (\revised{\figref{fig:fig1}C}), modeled as $\bl(g,Pe)=a_3(1+a_4\,Pe/g)^{-1/3}$ following a canonical L\'{e}v\^{e}que scaling \cite{Leal07}. Here $\uIcc$ is the centerline slip velocity for $clean$ finite-length gratings, found setting $\gMa=\gMaavg=0$ in \eqref{eq:flow_field}. This leads to $\uIcc=\uIcinf/[1+\qdinf(1-\gfx)]$, where $\uIcinf$ is the centerline slip velocity for infinite gratings, known from previous studies \cite{Philip_ZAMP_1972a,Teo2008-pe}. \revised{The effective slip length for clean gratings is $\lambda_e^\text{clean}=2\phi_x \qdinf/[3+\qdinf(3-4\gfx)]$. Although exact solutions of $\uIcinf$ and $\qdinf$ require numerical calculations, useful approximations are $\uIcinf \approx \{[(P/\pi)\cosh^{-1} (\sec (\pi \phi_z/2))]^{n_u} + 2^{n_u} \}^{1/n_u}$ and $q_d^\infty \approx \{[(3P/2\pi)\ln (\sec (\pi \phi_z/2))]^{n_q} + (3\phi_z)^{n_q} \}^{1/n_q}$, with $n_u=-1.46, n_q=-1.21$ (these become exact as $P \rightarrow 0$ or $P\rightarrow \infty$, see \SM\textit{ Flow field derivation)}.}

\revised{Equations~\eqref{eq:u_Ic}, \eqref{eq:lambdaE} link the loss of performance to the surfactant-induced stresses; these are enhanced by increasing the product of normalized concentration and Marangoni number $k\, Ma$, which expresses the ratio of Marangoni effects and viscous shear, for a unit of surfactant gradient. Marangoni stresses can be reduced by decreasing the surfactant gradient, achieved by either (i) increasing surfactant flux between the interface and the bulk (relative to advection), as captured by $Bi\, g^2 /(1+\delta\, Da)$, or (ii) increasing interfacial diffusivity (again, compared to advection), expressed by $1/Pe_I$.}

The scaling coefficients $a_1$, $a_2$, $a_3$ and $a_4$ in \eqref{eq:u_Ic},\revised{\eqref{eq:lambdaE}} are estimated by performing 155 simulations of the full governing equations, spanning a wide range of values in the dimensionless groups to ensure proper coverage of the parameter space (see \textit{Materials and Methods}). \figuref{fig:fit_agreement}A shows good agreement between the model \eqref{eq:u_Ic} and simulations across four orders of magnitude in the slip velocity, for $a_1\approx0.345$, $a_2\approx0.275$, $a_3\approx5.581$ and $a_4\approx3.922$, which are values of order one as expected for scaling coefficients. The simulations also corroborate the assumption $\gMa\approx\gMaavg$, as discussed in \SM\textit{ Finite-element simulations}.

Equipped with a 3D theory and a set of numerical simulation results, we aim to identify realistic combinations of the ten dimensionless parameters that maximize the drag reduction of SHSs.
Figs.~\ref{fig:lambda_vs_GF}B and C illustrate how $\lambda_e$ and $\uIc$ change with the spanwise gas fraction $\gfz$, for a fixed bulk concentration $\hcz=\SI{3e-4}{\mole\per\meter\tothe{3}}$ and $\hg=\SI{1}{\milli\meter}$, and for several values of surfactant solubility, in a micro-channel with $\hh=\SI{60}{\micro\meter}$.
\revised{Slip is maximized for surfactant with higher solubility, corresponding to lower $\hka/\hkd$. Incidentally, the results depend weakly on $\hka$ or $\hkd$ individually; although these parameters appear separately in Eqs.~\eqref{eq:surfactant}, \eqref{eq:lambdaE} through $Bi$ and $Da$, $Da$ is large if $\hh$ is larger than a few micrometer, such that the term $Bi/(1+\delta \,Da)\approx Bi/(\delta Da)$, which is a function of $\hka/\hkd$ (see also \SM\textit{ Discussion of the mobilization length}).}

Since the chemical properties of naturally occurring surfactants are virtually impossible to control in practice \cite{Hourlier-Fargette2018-tu}, we focus on geometrical parameters. We observe that $\gfz$ has a negligible impact on surfactant effects, even as $\gfz\to{1}$. Mathematically, $\gfz$ affects \equaref{eq:u_Ic} only through the surfactant-independent term $\uIcc(P,\gfz,\gfx)$, which is at most of order one. Equations~\equaref{eq:u_Ic} and \eqref{eq:lambdaE} reveal that slip is maximized by increasing the grating length $g$, which progressively overcomes surfactant effects, undergoing a transition of the form $\uIc \sim g^2$ and ultimately approaching the asymptotes $\uIc\to\uIcc$, $\lambda_e\to\lambda_e^\text{clean}$. This transition is challenging to simulate due to the large computational cost of long domains needed at large $g$. 

\begin{figure}[!t]
   \centering
   \captionsetup{farskip=0pt, nearskip=0pt}
   \includegraphics[width=\linewidth]{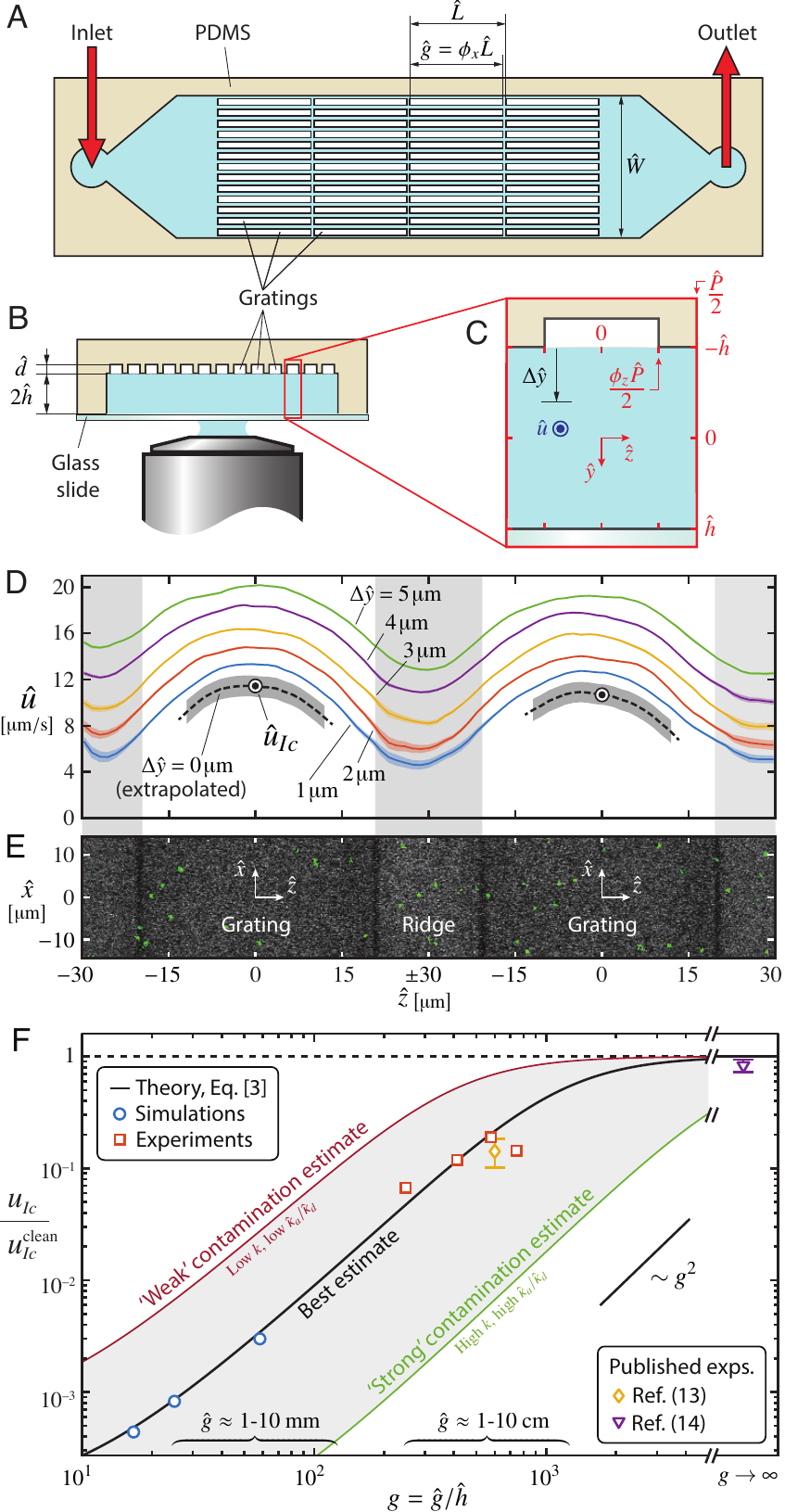}
   \caption{\label{fig:exp:top}\label{fig:exp:sec}\label{fig:exp:cell}\label{fig:exp:piv}\label{fig:exp:confocal}\label{fig:exp:results} (A) Top view and (B) cross section of the microfluidic channels used in the experiments. (C) Cross section of a unit cell, as defined in \revised{\figref{fig:fig1}B (note the inverted microscope setup).} The distance from the interface is $\Delta{\hat{y}}$. (D) Example of velocity profiles at different distances from the interface, for a grating length $\hat{g}=45\,$mm. The dashed line denotes the linearly extrapolated slip velocity, whereas the shadings show standard error (details in \SM\textit{ Experimental methods}). \revised{The bull's eyes mark the centerline slip velocity.} (E) Micrograph of the gratings shown in (D), with \uPIV~particles appearing in green. \revised{(F) Ratio between the actual centerline slip velocity and the surfactant-free value (i.e.~``clean''), $\uIc/\uIcc = \huIc/\huIc^\text{clean}$,} from our theory \eqref{eq:u_Ic}, simulations, and experiments, as well as prior experiments of \cite{peaudecerf17,Song2018-uw}. Ref.~\cite{Song2018-uw} used an annular grating in a rheometer, with an effectively infinite groove length. The theoretical prediction from our model \eqref{eq:u_Ic} and the simulations use a best estimate of the unknown surfactant parameters (see \SM\textit{ Estimate of surfactant parameters}), with the shaded region denoting a range of plausible levels of contamination within our estimates. The uncertainty in the present experimental data is smaller than the size of the symbols.}
   \label{fig:exp}
\end{figure}

\begin{figure*}[t!]
   \centering
   \captionsetup{farskip=0pt, nearskip=0pt}
   \subfloat{\includegraphics[]{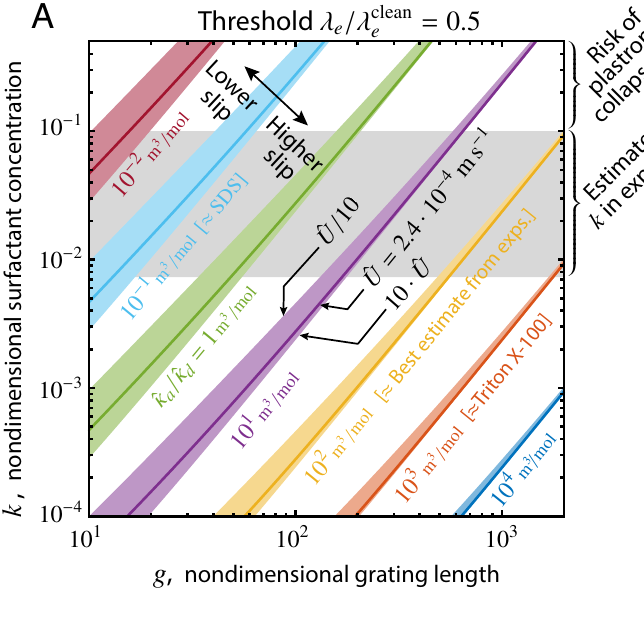}\label{fig:phase_space}}
   \subfloat{\includegraphics[]{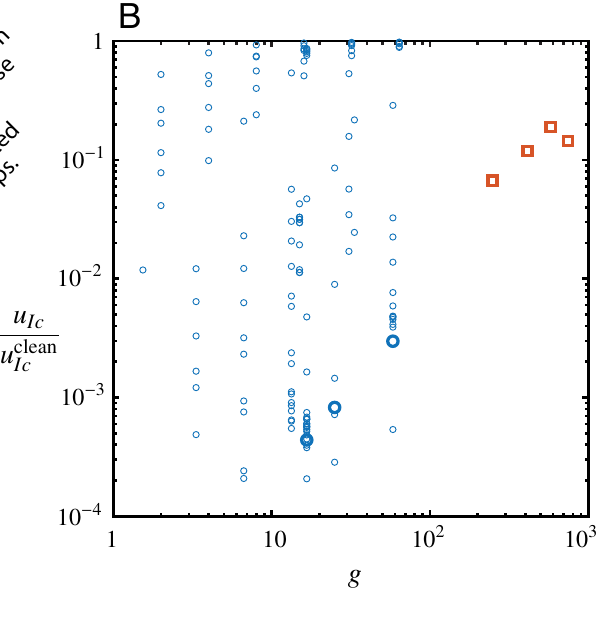}\label{fig:collapse_before}}
   \subfloat{\includegraphics[]{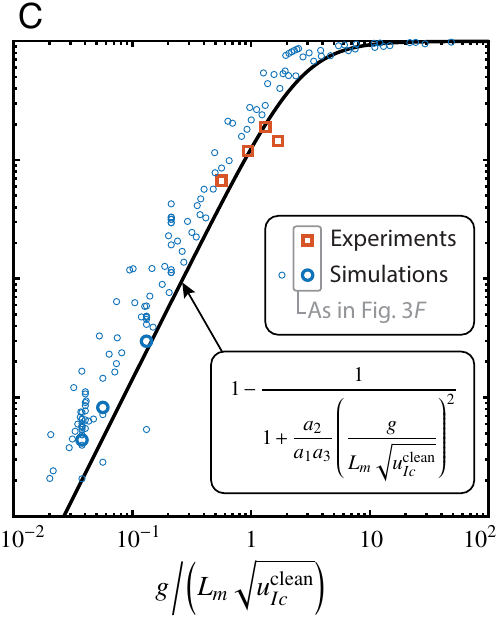}\label{fig:collapse_after}}
   \caption{(A) Lines (from our model) showing the normalized surfactant concentration $k$, as a function of grating length $g$, that yields a slip length that is 50\% of the ideal, `clean' value. Colors denote different surfactant solubilities, expressed by the ratio of the adsorption and desorption constants $\hka/\hkd$. Shaded bands show the weak effect of changing the flow velocity $\hU$  across two orders of magnitude. (B) Experimental and numerical data for the slip velocity $u_{Ic}$, plotted against $g$, as one varies surfactant properties, flow velocity, and SHS geometry, together characterized by ten dimensionless groups (including $g$). (C) Normalizing $g$ by the mobilization length $L_m$ approximately collapses the same data onto a single curve, governed by only one dimensionless group. Here the factor $(u_{Ic}^\text{clean})^{1/2}$ helps collapse data across a range of gas fractions, which give a wide range of values for $u_{Ic}^\text{clean}$. In practice the normalized clean interface velocity $u_{Ic}^\text{clean}$ is usually of $O(1)$, so this factor could be omitted in the normalization for $g$.  
   }
   \label{fig:mob_length_figs}
\end{figure*}

\section*{Experiments demonstrate effect of grating length}

To acquire data at large $g$ and test the prediction of a slip transition, we build microfluidic devices using polydimethylsiloxane (PDMS), 
as shown in \figref{fig:exp:top}A and \ref{fig:exp:sec}B (see \textit{Materials and Methods}). The channel upper wall consists of streamwise gratings of pitch $\hP=60\um$ and spanwise gas fraction $\gfz=2/3$. 
The channel half-height is $\hh=60\um$, and the depth of the grating trenches is $\hd=25\um$, enough to ensure a stable plastron throughout each experiment.
We test gratings with $\hg=15$, $25$, $35\text{ and }\SI{45}{m\meter}$, separated in the streamwise direction by solid ridges of length $\hL-\hg=\SI{20}{\micro\meter}$. 
We employ a confocal microscope and micro-particle image velocimetry (\uPIV) in a setup similar to the one in \cite{peaudecerf17}. A syringe pump provides a constant flow rate $\hat{Q}_{\text{TOT}}=\SI{1.152}{\micro\liter\per\minute}$. 
We use de-ionized water without any additives, since it has been established that the unavoidable amounts of surfactant naturally present in similar microfluidic settings are sufficient to induce significant stresses at the plastron \cite{Schaffel2016-mh,peaudecerf17,Song2018-uw}. The \uPIV~beads \revised{(visible in \figref{fig:exp:confocal}E)} are thoroughly pre-washed to remove their added surfactant \cite{Li20}, and we also follow a cleaning protocol for the syringes and tubing (see \textit{Materials and Methods}). 
The flow velocity is measured over two adjacent gratings, at several distances $\Delta\hat{y}$ from the interface, as defined in \figref{fig:exp:cell}C. Examples of velocity profiles are displayed in \figsref{fig:exp:piv}D, for $\hg=\SI{45}{m\meter}$. The flow over the solid ridges is consistent with the no-slip condition at the wall, whereas velocity increases noticeably over the gratings. These vertically spaced profiles around the grating centerline are extrapolated to obtain the slip velocity at the interface\revised{, shown by the black dashed lines in \figref{fig:exp:piv}D; the centerline slip velocity is marked by the bull's eyes in the figure.} \revised{The measured slip velocities $\hat{u}_{Ic}$  are only 6\% to 10\% of the values $\hat{u}_{Ic}^\text{clean}$} predicted by surfactant-free theories (Fig.~\ref{fig:exp:results}F), consistently with prior experiments \cite{Kim2012-iw,Bolognesi2014-vw,Schaffel2016-mh,peaudecerf17,Song2018-uw}.

Comparing quantitatively these experimental measurements to the predictions from our model requires assumptions on the type and amount of surfactant present in the channel. Although some parameter values are known and others can be accurately estimated, the normalized surfactant concentration $k$ and the kinetic rate adsorption and desorption constants $\hka$ and $\hkd$ can vary across a broad range. Nevertheless, it is possible to combine our model for the slip velocity \eqref{eq:u_Ic} with previous experimental results \cite{peaudecerf17} to obtain an estimate, as described in detail in \SM\textit{ Estimate of surfactant parameters}. We find approximate ranges for the normalized concentration $\SI{7.3e-3}{} \lesssim k \lesssim \SI{e-1}{}$ and for the ratio of constants $\SI{7.1e1}{\meter\tothe{3}\per\mole} \lesssim \hka/\hkd \lesssim \SI{1.8e3}{\meter\tothe{3}\per\mole}$.

Choosing the mid-range values $k=\SI{3.6e-2}{}$ and $\hka/\hkd=\SI{1.2e2}{\meter\tothe{3}\per\mole}$, our  predictions of the slip velocity show good agreement with our experimental data and with previous studies \cite{peaudecerf17,Song2018-uw}, as illustrated in \figuref{fig:exp:results}F. At small $g$, measuring the small slip velocity with high precision is challenging; we performed finite-element simulations with the same surfactant properties as in the experiments, shown by the blue circles in \figuref{fig:exp:results}F. These simulations are restricted to $g<60$ (approximately 3.5\,mm in practice), as computational cost increases with $g$. Simulations and experiments agree with \eqref{eq:u_Ic}, showing increased slip consistent with the theoretical prediction as $g$ increases \revised{and confirming the key impact of this geometric parameter in controlling drag reduction.} Furthermore, all our measurements are consistent with the range of $k$ and $\hka/\hkd$ that we estimated from
previous experiments in a different laboratory \cite{peaudecerf17}.

To assess practical implications for general microfluidic flows, we plot in \figuref{fig:phase_space} the combinations of $g$ and $k$ at which the slip length reaches 50\% of the surfactant-free value (i.e.~$\lame/\lamec = 0.5$, plotted with solid lines), for  surfactant solubility values $\hka/\hkd$ (plotted with different colors) and a range of representative velocities $\hU$ (shown by shaded colored bands around each line) found in small-scale applications.
The gray horizontal band shows the range $7 \cdot 10^{-3}\lesssim k\lesssim 0.1$ estimated for microfluidic experiments. At larger values $k > 0.1$ the risk of plastron collapse increases significantly due to capillary effects. 
Increasing or reducing the velocity $\hU$ by an order of magnitude has a relatively weak effect on the slip, as shown by the narrow bands around each line in \figref{fig:phase_space}.
In addition, \figref{fig:phase_space} shows that varying $\hka/\hkd$ leads to uniformly shifted contours, which remain approximately parallel. Together with the weak dependence on velocity, these results suggest that the 50\% threshold for slip could be expressed through a simpler underlying criterion.

\section*{Single lengthscale predicts interface mobility}
\revised{Figure~\ref{fig:exp:results}F shows that surfactant impairment is strongly dependent on the grating length $g$. We therefore define the {mobilization length} $L_m$ as the value of $g$ that gives significant slip, such that the local slip length on the interface is $\lambda_I(L_m) \sim 1$ (or in dimensional terms, $\hat{\lambda}_I(\hLm/\hat{h}) \sim \hat{h}$). We seek a scaling for $\lambda_I$ in terms of $g$, starting from the definition of $\lambda_I$ and using \eqref{eq:surfactant:marangoni}, }
\begin{equation}
\revised{
  \lambda_I = \frac{u_I}{\gamma} \sim \frac{u_I}{k Ma \dfrac{ \Delta \Gamma }{ g }}.}
  \label{eq:lambdaScaling}
\end{equation}
\revised{Note that $x$ is normalized by $\hL$, whereas $g$ is normalized by $\hh$, such that gradients along the interface occur over an $x$ length scale $\hg/\hL \sim \eps g$. Here $\Delta \Gamma$ is the order of magnitude of the change in $\Gamma$ along the interface, illustrated in Fig.~\ref{fig:fig1}D. More specifically, we assume that $c_I$ ranges from $1-\Delta c_I$ to $1+\Delta c_I$, and similarly $\Gamma$ is between $1 - \Delta \Gamma$ and $1 + \Delta \Gamma$, such that at some point near the middle of the interface $\Gamma \sim c\sim 1$. Scaling \eqref{eq:surfactant:kinetics},}
\begin{equation}
\revised{-\frac{\Delta c_I }{\delta} \sim Da \, (\Delta c_I - \Delta \Gamma),\quad \text{   i.e.   }\quad
\Delta \Gamma \sim \Delta c_I \left(\dfrac{1}{\delta\, Da }+1\right).}
\label{eq:fluxScaling}
\end{equation}

\revised{We find $\Delta c_I$ by scaling Eq.~\eqref{eq:surfactant:interface}, which represents a balance between advection, which acts to establish a surfactant gradient, and diffusion and kinetics, which act to extinguish the gradient. We consider changes along the $x$-direction, between the edge of the interface (where $u=0$) and a point near the middle (where $u\sim u_I$ and $\Gamma \sim 1$). Then $\partial_x(u \Gamma) \sim u_I / (\eps g)$ and $\partial_{xx}\Gamma \sim \Delta \Gamma/(\eps g)^2$. 
Writing the kinetic flux $Bi(c_I-\Gamma) = (Bi/Da)\,\partial_y{c}|_I$ and using \eqref{eq:fluxScaling} to eliminate $\Delta \Gamma$, Eq.~\eqref{eq:surfactant:interface} yields}
\begin{equation}
   \revised{ \frac{u_I}{g} \sim \left(Bi + \frac{1+\delta{Da}}{Pe_I{g^2}} \right)\frac{\Delta c_I }{\displaystyle \delta{Da}}.}
   \label{eq:advectionScalingFull}
\end{equation}
\revised{The second term in the parenthesis on the right hand side is negligible if diffusion is weak compared to kinetics, such that $g^2\gg (1+\delta Da)/(Bi \,Pe_I)$. Note that $\delta \sim 1$ unless the gratings are very short, and recall that $Da$ is large if $\hh$ is larger than a few micrometers, such that $(1+\delta \,Da) \sim Da $. In dimensional form, neglecting diffusion in \eqref{eq:advectionScalingFull} then requires $\hg^2\gg (\hLdmod)^2$, where}
\begin{equation}
     \revised{ \hLdmod \vcentcolon= \left(\hh\hLd\dfrac{\hDI}{\hD}\right)^{1/2},
      \label{eq:dep_length} }
\end{equation}
\revised{and $\hLd=\hka\hGamm/\hkd$ is the {depletion length}, which arises commonly in models of soluble surfactant \cite{Manikantan20}. We define $\hLdmod$ as a {modified} depletion length, representing the grating length above which interface diffusion is small compared to kinetics. Omitting the diffusive term in \eqref{eq:advectionScalingFull}, and combining with \eqref{eq:lambdaScaling} and \eqref{eq:fluxScaling} to eliminate $\Delta \Gamma$ and $\Delta c_I$, we find}
\begin{equation}
    \revised{    \lambda_I \sim  \dfrac{g^2\,Bi}{k\, Ma\, Da}.}
    \label{eq:lambdaScale}
\end{equation}
\revised{This scaling shows concisely that $\lambda_I$ depends on $g^2$, consistently with \eqref{eq:lambdaE} and with Fig.~\ref{fig:exp:results}F. 
If $\lambda_I \sim 1$ when $g \sim L_m$, \eqref{eq:lambdaScale} defines the mobilization length; in dimensional form, we find}
\begin{equation}
    \revised{\hLm \vcentcolon= \hat{h} \frac{\hat{\kappa}_a \hat{\Gamma}_m}{\hat{\kappa}_d} \left( \frac{\ns \hat{R} \hat{T} \hcz}{\hat{\mu} \hat{D}} \right)^{1/2}.}\label{eq:L_m}
\end{equation}
\revised{Marangoni stresses become negligible if $\hg^2 \gg \hLm^2$, as transport between the bulk and the interface suppresses the surfactant gradient that would otherwise be established by advection. For our experiments (Table SI), Eq.~\eqref{eq:L_m} predicts $\hLm \approx 4.3\,$cm, indicating that gratings must be at least several centimeters long to minimize Marangoni stresses, consistently with Fig.~\ref{fig:exp:results}F. Remarkably, $\hLm$ depends linearly on $\hka/\hkd$ but only on the square-root of $\hcz$, showing higher sensitivity to the type of surfactant than to its concentration.}

\revised{For small-scale applications, and using properties of known surfactants, we find that $\hLdmod$ is much smaller than $\hLm$; for example, $\hLdmod \approx \SI{170}{\micro \metre}$ in our experiments (\textit{Table~SI} and \SM\textit{ Discussion of the mobilization length}). Therefore the mobilization length $\hLm$ is the key length scale that determines the slip of a given SHS.}

\revised{Our analysis suggests that it should be possible to write simplified expressions for $\uIc$ and $\lambda_e$ in terms of $g/L_m$. With the approximation $(1+\delta{Da})\approx {Da}$ (as discussed above) and assuming $\hLm\gg\hLdmod$ (such that the terms $1/Pe_I$ are negligible), Eqs.~\eqref{eq:u_Ic} and \eqref{eq:lambdaE} simplify to the one-parameter curves}
\begin{align}
   \dfrac{\uIc}{\uIcc} &= 1 - \dfrac{1}{1+\dfrac{a_2}{a_1a_3}\left(\dfrac{g}{\Lm\sqrt{\uIcc}}\right)^2} ,
  \label{eq:universal_curve_u}\\
 \revised{ \dfrac{\lambda_e}{\lambda_e^\text{clean}}} &= \revised{ 1 - \dfrac{1}{1+\dfrac{a_2}{a_1a_3}\left(\dfrac{g}{\Lm\sqrt{\uIcc \left(1+\frac{\lambda_e^\text{clean}}{2} \right)}}\right)^2}. }
 \label{eq:universal_curve_lambda}
\end{align}
\revised{We re-examine the 155 simulations shown in \figref{fig:fit_agreement}A, and set aside 24 cases that are not achievable in reality (e.g.~involving unphysically small diffusivities). Figure~\ref{fig:collapse_before} shows the relative slip $\uIc/\uIcc$ versus $g$ for the remaining 131 simulations and for our experiments.
These data span different surfactant properties, grating geometries and flow velocities, and exhibit large scatter, illustrating how $g$ alone is insufficient to predict slip. However, when $g$ is normalized by the mobilization length $\Lm$, the same data collapse near the one-parameter curve given by \eqref{eq:universal_curve_u}, as shown in \figref{fig:mob_length_figs}C. }

\section*{Discussion and outlook}
Regarding the surfactant type inherent to our experiments, Eqs.~\eqref{eq:L_m} and \eqref{eq:universal_curve_u} suggest a surfactant with large $\hka/\hkd$, implying low solubility. This is consistent with previous findings that PDMS used in microfluidic channels (including our experiments) releases uncrosslinked oligomer chains \cite{Kim:06,Eddington:06,Wong:20}, which are surface active \cite{Hourlier-Fargette17,Hourlier-Fargette2018-tu}, and which have also been detected in solution \cite{Regehr09,carter:20}. The mass fractions reported in \cite{carter:20}, in combination with the oligomer chain lengths found in \cite{Regehr09}, lead to concentrations $\hcz\sim{O}(10^{-4}-10^{-2})$ \SI{}{\mole\per\meter\tothe{3}}, compatible with our estimates. Incidentally, in other contexts, PDMS has sometimes been approximately modeled as insoluble \cite{lee:91,bergeron:96}; under this assumption, our theory yields $\uIc=\uIcc/(1+a_\text{ins}\,Ma_\text{ins}\,\uIcc)$ (see \SM\textit{ Scaling theory for surfactant transport}), where $a_\text{ins}$ is a scaling coefficient, $Ma_\text{ins}=\ns\hR\hT\hGamz\hh/(\visc\hDI)$ is a Marangoni number, and $\hGamz$ is the average interfacial surfactant concentration. Note that this expression for $\uIc$ does not depend on $g$, inconsistently with the experimental results in \figref{fig:exp:results}F. This highlights the importance of including solubility in models of surfactant dynamics on SHSs.

The results described here provide insights about the slip and drag of superhydrophobic surfaces in realistic conditions. Our theory for slender, {finite} gratings (which are widely used) enables comparisons with experiments, where inherent surfactants must be accounted for. The hydrodynamic component of the model from \eqqref{eq:flow_field} can also quantify SHS performance in the presence of general, nonuniform shear stresses at the air-water interface, thus circumventing the need for computationally expensive simulations. In addition, we have shown that a single mobilization length scale arising from the theory can serve as a guide in the design of SHS textures that mitigate surfactant effects. 
This laminar theory is also a stepping stone towards predicting surfactant effects in turbulent flow. As a first approximation, the mobilization length could be estimated by replacing the channel half-height $\hat{h}$ in \eqref{eq:L_m} with the thickness of the viscous sublayer. Finally, since $\hLm$ depends primarily on surfactant properties and on the shear length scale $\hh$, we may also expect that the concept of mobilization length, derived here for streamwise gratings, could qualitatively apply to other SHS textures.

\matmethods{%
\subsection*{Finite-element simulations} We solved the full governing equations and boundary conditions (in dimensional form, detailed in \SM\textit{ Governing equations}) in three dimensions using COMSOL Multiphysics 5.5. We performed a total of 155 simulations using different grating geometries, flow velocities and surfactant properties in order to span a large portion of the parameter space characterized by the ten dimensionless numbers of the problem. The domain was one half of the SHS unit cell depicted in \figref{fig:fig1}B, with $\hz$ between $\hz = 0$ and $\hz = \hP/2$ due to the spanwise symmetry of the solution. The volume is meshed with tetrahedral elements, with the finest ones (with a minimum element size of \SI{1.5e-9}{\meter}) around the upstream and downstream edges of the interface $\hx = \pm\gfx\hL/2$ (see \SM\textit{ Finite-element simulations}). We used the Creeping Flow module for the flow field, the Dilute Species Transport module for the transport of bulk surfactant, and the transport of interfacial surfactant was implemented through a General Form Boundary PDE. The Marangoni boundary conditions were enforced through a Weak Contribution constraint, as was the condition that fixed the mean bulk concentration to be $\hcz$. The system of nonlinear equations was solved through a Newton-type iterative method using the PARDISO direct solver for the linear system at each iteration. All simulations satisfied a relative tolerance for convergence of \SI{e-5}{}. We used linear elements for the pressure, bulk concentration and interfacial concentration, and either linear or quadratic elements for the velocity field, depending on the computational demands of each simulation.

\subsection*{Microchannel fabrication} Microfluidic channels with an array of parallel SHS gratings on their ceiling (\figsref{fig:exp:top}A and \ref{fig:exp:sec}B) were built by casting PDMS (Sylgard 184) over a mold obtained by two-layer photolithography. The photoresist used was SU-8 (Microchem SU-8 3025 and Microchem SU-8 3050). The chips were bonded to 0.1\,mm-thick glass coverslips (Bellco Glass 1916-25075) through untreated adhesion. Every coverslip was washed with isopropyl alcohol, then with \SI{18}{\mega\ohm\centi\meter} DI water, and finally dried with nitrogen before the microfluidic chip was attached. \revised{The static contact angle of water droplets was measured to be higher than 100\textdegree\, over samples of smooth PDMS, and to increase further over samples of textured PDMS. This is consistent with previous measurements for untreated PDMS \cite{Miranda:22}, and demonstrates the superhydrophobicity of the substrate. The total width of the microfluidic channel was set to $\hW=\SI{2}{\milli\meter}$ (see \figref{fig:exp:top}A), to ensure an approximately periodic flow in $\hz$ over the gratings far away from lateral walls (given that $\hW\gg\hh$).}

\subsection*{Experimental setup} A glass syringe (Hamilton Gastight) was filled with particle-seeded DI water, which was driven through the microchannels using a syringe pump (KD Legato 111). We used the barrel of a plastic syringe (BD Luer-Lok) as an outlet reservoir open to the room, to impose atmospheric pressure at the end of the circuit. The height of this reservoir was adjusted with a vertical translation stage (Thorlabs VAP10) at the beginning of each experiment to ensure that the plastron at each grating remained approximately flat, by controlling the average pressure in the microchannel. The microchannel was connected to the syringe and reservoir through plastic tubing (Tygon S3). All circuit elements were thoroughly pre-washed with \SI{18}{\mega\ohm\centi\meter} DI water, following a protocol described in \SM\textit{ Experimental Methods}.

\subsection*{Confocal microscopy} The tracer particles (ThermoFisher FluoSpheres carboxylate \SI{0.5}{\micro\meter} diameter) were washed using a centrifuge (Eppendorf 5418) to separate them from the buffer solution, which was then discarded and replenished with \SI{18}{\mega\ohm\centi\meter} DI water. This process was repeated three times before each experiment to essentially eliminate surfactant contamination from the particle solution. The flow was observed with a confocal microscope (Leica SP8 Resonant Scanning), using a 40X water objective (as in \figref{fig:exp:sec}B). The microfluidic device was enclosed in a stage top chamber (Okolab H101-K-FRAME) with a controlled temperature set to $\hT=\SI{296}{\kelvin}$. Using the bright field imaging of the microscope (\figref{fig:exp:confocal}E), we focused on two adjacent gratings. We avoided imaging the five gratings closest to each lateral side wall of the channel to prevent effects related to the loss of periodicity. The fluorescence imaging of the microscope (superimposed on the image in \figref{fig:exp:confocal}E) revealed the positions of the tracer beads in each snapshot, which we obtained at a rate of between 20 and 28 frames per second. All the data were taken at the center of the grating in the streamwise direction (i.e.~$\hx\approx{0}$) and at several distinct $\hy$-planes close to the interface (see \figref{fig:exp:piv}D).

\subsection*{Image analysis and micro-PIV} 
The \uPIV~analysis was performed with the open-source MATLAB toolbox PIVlab \cite{PIVlab:14}, using an acquisition window of approximately \SI{125}{\micro\meter}$\times$\SI{125}{\micro\meter}. The velocity field obtained for a given window was averaged in time and along the streamwise $\hx$ direction to
obtain the spanwise velocity profiles depicted in \figref{fig:exp:piv}D, for different distances away from the interface. To extract the centerline slip velocity $\huIc$ (at $\hy=-\hh$ and $\hz=0$), we performed a linear least-squares fit using data from between three to five $\hy$-planes. We only used data in a neighborhood of the grating center $\hz=0$ (see \figref{fig:exp:piv}D), since velocity profiles at $\hy=-\hh$ were not smooth due to the transitions between the interfaces and the solid ridges at $\hz=\pm\gfz\hP/2$. The uncertainty for $\huIc$ was calculated by accounting for how uncertainties in the velocity measurements and in the $\hy$-coordinate of the interface ($\pm\SI{1}{\micro\meter}$) propagated through the fitting procedure.

}

\showmatmethods{} 

\acknow{We thank David Bothman, Benjamin Lopez and Rachel Schoeppner for technical assistance. This work is supported by NSF CAREER 2048234, ARO MURI W911NF-17-1-0306, ONR MURI N00014-17-1-2676, the California NanoSystems Institute, and the European Union’s Horizon 2020 research and innovation programme under Marie Skłodowska-Curie grant agreement No.~798411. We acknowledge the use of the NRI-MCDB Microscopy Facility supported by NSF MRI grant DBI-1625770. A portion of this work was performed in the Microfluidics Laboratory within the California NanoSystems Institute, supported by the University of California Santa Barbara, and the University of California, Office of the President.}

\showacknow{} 

\bibliography{SHS3D_biblio}

\end{document}



\maketitle


\section*{Governing equations}\label{si:sec:gov_eqns}

We consider a steady fluid flow at low Reynolds number, within the unit cell depicted in Fig.~1A,~B. The three-dimensional velocity field is denoted by $\huvec=\hu\,\ex+\hv\,\ey+\hw\,\ez$, where $\ex$, $\ey$ and $\ez$ are unit vectors in the streamwise, wall-normal and spanwise directions (see Fig.~1C). The scalar fields $\hp$ and $\hc$ represent the pressure and the bulk surfactant concentration, respectively. The governing equations describing the conservation of mass, momentum, and surfactant in the bulk fluid are, in dimensional form,
\begin{subequations} \label{si:eq:dim}
    \begin{align} 
         \pd{\hu}{\hx}{} + \pd{\hv}{\hy}{} + \pd{\hw}{\hz}{}                        &= 0,                  \label{si:eqdim:cont} \\
         \visc\left( \pd{\hu}{\hx}{2} + \pd{\hu}{\hy}{2} + \pd{\hu}{\hz}{2} \right) &= \pd{\hp}{\hx}{},    \label{si:eqdim:u}    \\
         \visc\left( \pd{\hv}{\hx}{2} + \pd{\hv}{\hy}{2} + \pd{\hv}{\hz}{2} \right) &= \pd{\hp}{\hy}{},    \label{si:eqdim:v}    \\
         \visc\left( \pd{\hw}{\hx}{2} + \pd{\hw}{\hy}{2} + \pd{\hw}{\hz}{2} \right) &= \pd{\hp}{\hz}{},    \label{si:eqdim:w}    \\
         \hu\pd{\hc}{\hx}{} + \hv\pd{\hc}{\hy}{} + \hw\pd{\hc}{\hz}{}               &= \hD  \left( \pd{\hc}{\hx}{2} + \pd{\hc}{\hy}{2} + \pd{\hc}{\hz}{2} \right).                      \label{si:eqdim:c}    
    \end{align}
    At the interface, the interfacial surfactant concentration $\hGam$ follows a conservation law. An adsorption--desorption flux couples $\hGam$ to the bulk concentration. Marangoni boundary conditions link the interfacial shear stress to the concentration gradient. The corresponding equations, defined only at the air--water interface, read
    \begin{align} 
         \pd{(\hu_I\hGam)}{\hx}{} + \pd{(\hw_I\hGam)}{\hz}{} &= \hDI \left( \pd{\hGam}{\hx}{2} + \pd{\hGam}{\hz}{2} \right) + \hkin,   \label{si:eqdim:Gam}  \\
         \hD\left.\pd{\hc}{\hy}{}\right|_I                   &= \hkin,                                                                 \label{si:eqdim:ads-des}  \\
         \visc\left.\pd{\hu}{\hy}{}\right|_I                 &= \hnonlin\pd{\hGam}{\hx}{},                                             \label{si:eqdim:Ma_u} \\
         \visc\left.\pd{\hw}{\hy}{}\right|_I                 &= \hnonlin\pd{\hGam}{\hz}{},                                             \label{si:eqdim:Ma_w}
    \end{align}
    where $\hnonlin$ is a possibly nonlinear term quantifying the dependence of the surface tension with $\hGam$, and depends on the specific model of equilibrium isotherm chosen \cite{Manikantan20}. The term $\hkin$ represents the adsorption-desorption kinetics, and must be compatible with the choice of isotherm. Here, we use a model derived from the Frumkin isotherm \cite{Chang_Franses_1995,prosser01}, which leads to
    \begin{align} 
         \hkin    &= \hka\hc_I(\hGamm-\hGam) - \hkd\hGam{e^{A\hGam/\hGamm}},                      \label{si:eqdim:kin}    \\
         \hnonlin &= \ns\hR\hT\left(\dfrac{\hGamm}{\hGamm-\hGam}+A\dfrac{\hGam}{\hGamm}\right).   \label{si:eqdim:nonlin}
    \end{align}
    The above equations are complemented with the imposition of a mean background level of bulk concentration
    \begin{align} 
         \dfrac{1}{2\hh\hP\hL}\int_{-\hP/2}^{\hP/2}\int_{-\hh}^{\hh}\int_{-\hL/2}^{\hL/2}\hc\,\mathrm{d}\hx\,\mathrm{d}\hy\,\mathrm{d}\hz &= \hcz, \label{si:eqdim:c_zero}
     \end{align}
     as well as with streamwise and spanwise periodicity conditions for variables defined in the bulk fluid, 
     \begin{align}
         \huvec(\hxvec)    &= \huvec(\hxvec + \alpha\hL\ex + \beta\hP\ez)              \,\,\,\,\text{for any integers $\alpha$, $\beta$,} \label{si:eqdim:period_u}      \\
         \hc(\hxvec)       &= \hc(\hxvec + \alpha\hL\ex + \beta\hP\ez)                 \,\,\,\,\text{for any integers $\alpha$, $\beta$,} \label{si:eqdim:period_c} 
     \end{align}
     which in the case of the pressure also includes a mean pressure drop such that
     \begin{align}
         \hp(\hxvec)       &= \hp(\hxvec + \alpha\hL\ex + \beta\hP\ez) + \alpha\hG\hL  \,\,\,\,\text{for any integers $\alpha$, $\beta$,} \label{si:eqdim:period_p}
     \end{align}
     and where $\hxvec=\hx\,\ex+\hy\,\ey+\hz\,\ez$ is the position vector. The remaining equations are the boundary conditions
     \begin{align}
         \huvec            &= \bm{0}  \,\,\,\,\text{on all solid surfaces (no slip and no penetration),} \label{si:eqdim:no_slip}                        \\
         \hv               &= 0       \,\,\,\,\text{on the air--water interface (no penetration),} \label{si:eqdim:no_pen}                       \\
         \pd{\hc}{\hy}{}   &= 0       \,\,\,\,\text{on all solid surfaces (no flux),} \label{si:eqdim:no_fl_c}                        \\
         \pd{\hGam}{\hx}{} &= 0       \,\,\,\,\text{at $\hx=\pm\gfx\hL/2$ when $|\hz|\leq\gfz\hP/2$ (no flux),} \label{si:eqdim:no_fl_Gx} \\
         \pd{\hGam}{\hz}{} &= 0       \,\,\,\,\text{at $\hz=\pm\gfz\hP/2$ when $|\hx|\leq\gfx\hL/2$ (no flux).} \label{si:eqdim:no_fl_Gz}
    \end{align}
\end{subequations}

We normalize Eqs.~\ref{si:eqdim:cont}-\ref{si:eqdim:no_fl_Gz} following
\begin{equation}  \label{si:nondim}
    \begin{gathered}
         x = \hx/\hL,\quad y = \hy/\hh = \hy/(\eps\hL),\quad z = \hz/\hh =\hz/(\eps\hL) \\
         u = \hu/\hU,\quad v = \hv/(\eps\hU),\quad w = \hw/(\eps\hU),\quad p = \hp/(\hG\hL) \\
         c = \hc/\hcz,\quad \Gam = \hGam/\hGamz,
    \end{gathered}
\end{equation}
where $\hU = \hh^2\hG/\visc$ and $\hGamz = \hka\hcz\hGamm/\hkd$ are the natural scales for the velocity and the interfacial surfactant. Applying this normalization to Eqs.~\ref{si:eqdim:cont}-\ref{si:eqdim:nonlin} results in
\begin{subequations} \label{si:eqnondim}
    \begin{align} 
         \pd{u}{x}{} + \pd{v}{y}{} + \pd{w}{z}{}                               &= 0,              \label{si:eqnondim:cont} \\
         \left( \eps^2\pd{u}{x}{2} + \pd{u}{y}{2} + \pd{u}{z}{2} \right)       &= \pd{p}{x}{},    \label{si:eqnondim:u}    \\
         \eps^2\left( \eps^2\pd{v}{x}{2} + \pd{v}{y}{2} + \pd{v}{z}{2} \right) &= \pd{p}{y}{},    \label{si:eqnondim:v}    \\
         \eps^2\left( \eps^2\pd{w}{x}{2} + \pd{w}{y}{2} + \pd{w}{z}{2} \right) &= \pd{p}{z}{},    \label{si:eqnondim:w}    \\
         u\pd{c}{x}{} + v\pd{c}{y}{} + w\pd{c}{z}{}                            &= \dfrac{1}{\eps{Pe}}\left( \eps^2\pd{c}{x}{2} + \pd{c}{y}{2} + \pd{c}{z}{2} \right).                      \label{si:eqnondim:c}    \\
         \pd{(u_I\Gam)}{x}{} + \pd{(w_I\Gam)}{z}{}                         &= \dfrac{1}{\eps{Pe_I}} \left( \eps^2\pd{\Gam}{x}{2} + \pd{\Gam}{z}{2} \right) + \dfrac{Bi}{\eps}\kin,   \label{si:eqnondim:Gam}  \\
         \left.\pd{c}{y}{}\right|_I                                            &= Da\,\kin,                                                                 \label{si:eqnondim:ads-des}  \\
         \left.\pd{u}{y}{}\right|_I                 &= \eps{k}Ma\,\nonlin\pd{\Gam}{x}{},                                             \label{si:eqnondim:Ma_u} \\
         \eps^2\left.\pd{w}{y}{}\right|_I                 &= \eps{k}Ma\,\nonlin\pd{\Gam}{z}{},                                             \label{si:eqnondim:Ma_w} \\
         \kin    &= c_I(1-k\Gam) - \Gam{e^{kA\Gam}},                      \label{si:eqnondim:kin}    \\
         \nonlin &= \left(\dfrac{1}{1-k\Gam}+kA\Gam\right).   \label{si:eqnondim:nonlin} 
     \end{align}
\end{subequations}

\section*{Problem parameters}\label{si:sec:param}

The parameters appearing in Eqs.~\ref{si:eqdim:cont}-\ref{si:eqnondim:nonlin} are defined in Tables \ref{si:tab:dim_numbers} and \ref{si:tab:nondim_numbers}, together with the values that they take in our experiments. Since, as explained in the main text, the surfactant type and concentration in the liquid are unknown, only an estimate can be obtained in some cases (see Section \textit{Estimate of surfactant parameters} for details). We choose $g=\hg/\hh$, $P=\hP/\hh$, $\gfx$ and $\gfz$ as the four independent geometric parameters of the problem, noting that $\eps$ can then be obtained as $\eps=\gfx/g$. Tables \ref{si:tab:dim_numbers} and \ref{si:tab:nondim_numbers} also include the values of the length scales $\hLdmod$ and $\hLm$, defined in Eqs.~10 and 9 of the main text.

\begin{table}[!h]
    \centering
        \caption{Parameters appearing in the dimensional Eqs.~\ref{si:eqdim:cont}--\ref{si:nondim} and defining the flow geometry (Fig.~1), alongside with their values in the simulations and experiments. The symbol $\ddagger$ indicates \revised{quantities whose order of magnitude does not change appreciably across surfactants; the values are for sodium dodecyl sulfate (SDS), which is well characterized \cite{Chang_Franses_1995}.} The symbol $\dagger$ denotes values that have been estimated by combining our theory and the experimental results in \cite{peaudecerf17} (see Section \textit{Estimate of surfactant parameters} for details).}
    \label{si:tab:dim_numbers}
    \begin{tabular}{ |c|c|c||c||c| }
        \hline
        Quantity & Symbol & Units & \begin{tabular}{@{}c@{}} Value (or best estimate)\\ in experiments\end{tabular} \\
        \hline
        \hline
        Background bulk concentration                          &  $\hcz$              & \SI{}{\mol\per\meter\tothe{3}}            & \SI{3e-4}{} $\dagger$                     \\
        Adsorption rate constant                               &  $\hka$              & \SI{}{\meter\tothe{3}\per\mol\per\second} & \SI{8.95e1}{} $\dagger$ \\
        Desorption rate constant                               &  $\hkd$              & \SI{}{\per\second}                        & \SI{7.5e-1}{} $\dagger$ \\
        Maximum packing concentration                          &  $\hGamm$            & \SI{}{\mole\per\square\meter}             & \SI{3.9e-6}{} $\ddagger$             \\
        Bulk surfactant diffusivity                            &  $\hD$               & \SI{}{\square\meter\per\second}           & \SI{7e-10}{} $\ddagger$              \\
        Interface surfactant diffusivity                       &  $\hDI$              & \SI{}{\square\meter\per\second}           & \SI{7e-10}{} $\ddagger$              \\
        Salinity parameter                                     &  $\ns$               & -                                         & $2$ $\ddagger$                       \\
        Frumkin interaction coefficient                                &  $A$                 & -                                         & $-2.4$ $\ddagger$                    \\
        Ideal gas constant                                     &  $\hR$               & \SI{}{\joule\per\mole\per\kelvin}         & $8.314$                              \\
        Temperature                                            &  $\hT$               & \SI{}{\kelvin}                            & $296$                                \\
        Dynamic viscosity                                      &  $\visc$             & \SI{}{\kilogram\per\meter\per\second}     & \SI{8.9e-4}{}                        \\
        Velocity scale                                         &  $\hU$               & \SI{}{\meter\per\second}                  & \SI{2.4e-4}{}                        \\
        Channel half height (see Fig.~1)   &  $\hh$               & \SI{}{\meter}                             & $\SI{6e-5}{}\pm\SI{3e-6}{}$          \\
        Pitch (see Fig.~1)                 &  $\hP$               & \SI{}{\meter}                             & \SI{6e-5}{}                          \\
        Grating width (see Fig.~1)         &  $\gfz\hP$           & \SI{}{\meter}                             & \SI{4e-5}{}                          \\
        Grating length (see Fig.~1)        &  $\hg$               & \SI{}{\meter}                             & $(1.5,2.5,3.5,4.5)\cdot\SI{e-2}{}$   \\
Ridge size in $x$ (see Fig.~1)             &  $\hL-\hg$  & \SI{}{\meter}                             & \SI{2e-5}{}                          \\
        \hline
        \hline
        Depletion length (see Eq.~8)                      &  $\hLd\revised{ = \hka\hGamm/\hkd}$              & \SI{}{\meter}                             & \SI{4.7e-4}{}                        \\
        \quad Modified depletion length (see Eq.~8) \quad &  $\hLdmod\revised{= \left(\hh\hDI\hka\hGamm/(\hD\hkd)\right)^{1/2}}$           & \SI{}{\meter}                             & \SI{1.7e-4}{}                       \\
        Mobilization length (see Eq.~10)                      &  $\hLm\revised{=(\hh\hka\hGamm/\hkd) \left(\ns\hR\hT\hcz/(\visc\hD)\right)^{1/2}}$              & \SI{}{\meter}                             & \SI{4.3e-2}{}                        \\
        \hline
    \end{tabular}

\end{table}

\vspace{15pt}
    
\begin{table}[!h]
    \centering
        \caption{Dimensionless numbers governing the full problem.}
    \label{si:tab:nondim_numbers}
    \renewcommand{\arraystretch}{1.3} 
    \begin{tabular}{ |c|c||c||c|  }
        \hline
        Dimensionless group & Definition & \begin{tabular}{@{}c@{}} Range\\ in simulations\end{tabular} & \begin{tabular}{@{}c@{}} Value (or best estimate)\\ in experiments\end{tabular} \\
        \hline
        \hline
        Normalized concentration    & \quad $k=\hka\hcz/\hkd = \hGamz/\hGamm$ \quad & \SI{2.7e-5}{} -- \SI{5.4e-2}{}  & \SI{4e-2}{}               \\
        Marangoni number            & $Ma=\ns\hR\hT\hGamm/(\visc\hU)$   & \SI{3.1e3}{}  -- \SI{2.3e7}{}   & \SI{9e4}{}                \\
        P{\'e}clet number           & $Pe=\hh\hU/\hD$                   & \SI{1.5e-2}{} -- \SI{1.2e5}{}   & \SI{2e1}{}                \\
        Interface P{\'e}clet number & $Pe_I=\hh\hU/\hDI$                & \SI{1.7e-1}{} -- \SI{6e2}{}     & \SI{2e1}{}                \\
        Biot number                 & $Bi=\hh\hkd/\hU$                  & \SI{8.6e-3}{} -- \SI{2.5e2}{}   & \SI{2e-1}{}                \\
        Damk{\"o}hler number        & $Da=\hh\hka\hGamm/\hD$            & \SI{2.5e1}{}  -- \SI{6.4e3}{}   & \SI{3e1}{}                \\
        Normalized grating length   & $g=\hg/\hh=\gfx/\eps$             & \SI{1.54}{}   -- \SI{58.33}{}   & \SI{2.5e2}{} -- \SI{7.5e2}{} \\
        Normalized pitch            & $P=\hP/\hh$                       & $0.92          - 2$              & $1  $                          \\
        Streamwise gas fraction     & $\gfx$                            & $0.833         - 0.994$          & $0.9986 - 0.9995$             \\
        Spanwise gas fraction       & $\gfz$                            & $0.667         - 0.980$          & $0.667$                          \\
        \hline
        \hline
        Normalized depletion length       & $L_d = \hLd/\hh = Da/(Pe{Bi})$                            & \SI{3.4e-3}{} -- \SI{2.7e4}{}          & \SI{7.8e0}{}                          \\
        \quad Normalized modified depletion length \quad       & $L_d^\text{mod} = \hLdmod/\hh\revised{  = \left(Da/(Pe_I{Bi})\right)^{1/2}}$                            & \SI{1.2e-1}{} -- \SI{4.9e1}{}          & \SI{2.8e0}{}                          \\
        \quad Normalized mobilization length \quad       & $\Lm = \hLm/\hh\revised{ = (k\,Ma\,Da/Bi)^{1/2}}$                            & \SI{2e0}{} -- \SI{1.1e3}{}          & \SI{7e2}{}                          \\
        \hline
    \end{tabular}

\end{table}

\section*{Flow field derivation} \label{si:sec:th_flow}
\subsection*{Assumption of a spanwise constant interface shear stress}  \label{si:sec:th_flow_spanwise}
Note that, although  $\eps\ll{1}$ and $k\ll{1}$ in the conditions considered in our study (see Section~\textit{Scaling theory for surfactant transport}), the product $\eps{k}{Ma}$ appearing in \eqref{si:eqnondim:Ma_u} and \eqref{si:eqnondim:Ma_w} is typically not small, since the Marangoni number is expected to be large, i.e.~$Ma\gg{1}$ (see estimates in \tabref{si:tab:nondim_numbers}) and the term $\nonlin\approx{1}$ as long as $k$ and $k|A|$ remain small. In fact, \equaref{si:eqnondim:Ma_u} implies that only when $\eps{k}{Ma}\gtrsim{1}$ the Marangoni stresses at the interface are non-negligible, as it is observed experimentally \cite{Bolognesi2014-vw, Schaffel2016-mh, peaudecerf17, Song2018-uw}. Since $\eps\ll{1}$, it is possible to assume that $\eps{k}{Ma}\gtrsim{1}\gg{\eps^2}$, and in that case it follows from \eqref{si:eqnondim:Ma_w} that $\partial_{z}\Gam\approx{0}$ at leading order in $\eps$. As detailed in the main text, the asymptotic expansion leading to \eqref{si:eqnondim:Ma_u} and \eqref{si:eqnondim:Ma_w} is singular, and thus the approximation $\partial_{z}\Gam\approx{0}$ is valid only in regions far from the upstream and downstream stagnation points, i.e. for $|x\pm\gfx/2|\gg\eps$. Indeed, our finite-element simulations of the full problem confirm that this approximation remains valid in all the regimes considered (as illustrated in  \figref{si:fig:sim_intf}, showing the contours of $\Gamma$). The Marangoni shear $\gMa=\left.\partial_y{u}\right|_I$ is thus also assumed to be independent of $z$ and only dependent on $x$, following \eqref{si:eqnondim:Ma_u}.  
\subsection*{Velocity field}
At leading order in the small parameter $\eps$, Eqs.~\ref{si:eqnondim:u}-\ref{si:eqnondim:w} representing the flow field are
\begin{subequations} \label{si:eq:flow_eqns}
    \begin{align} 
         \pd{u}{y}{2} + \pd{u}{z}{2} &= \pd{p}{x}{} , \label{si:eq:gov_eq_u} \\
         \pd{p}{y}{}  = \pd{p}{z}{}  &= 0           . \label{si:eq:gov_eq_p}
    \end{align}
\end{subequations}

It is clear from \eqref{si:eq:gov_eq_p} that $p$, and thus also $\partial_x{p}$, only depends on $x$. Since the solution $u$ depends on $x$ only through the right-hand-side of \eqref{si:eq:gov_eq_u}, we pose a piecewise solution
\begin{equation} \label{si:eq:piecewise}
u(x,y,z) =  
    \begin{cases}
        u_1(x,y,z) &\quad \text{  if } |x| < \gfx/2, \\
        u_2(x,y,z) & \quad\text{  if } \gfx/2 \leq |x| \leq 1/2. \\
    \end{cases}
\end{equation}

Taking into account the boundary conditions, the function $u_1$ satisfies the mixed boundary-value problem given by \eqref{si:eq:gov_eq_u} and the boundary conditions
\begin{equation} \label{si:eq:u1}
    \begin{aligned}
         u_1 &= 0                \quad\text{ if } y=1 \text{ or if } y=-1 \text{ and } |z|\geq\gfz{P} , \\
         \pd{u_1}{y}{} &= \gMa \quad \text{ if } y=-1 \text{ and } |z|<\gfz{P} .
    \end{aligned}
\end{equation}
We then introduce the Poiseuille profile $u_P(y) = (1-y^2)/2$ and, by virtue of the linearity of the problem, decompose the solution following $u_1 = -\left[\partial_x{p}(x)\right]u_P(y) - \left[\gMa+\partial_x{p}(x)\right]u_d^\infty$. The resulting problem for $u_d^\infty$ is homogeneous, yielding
\begin{equation} \label{si:eq:udinf}
    \begin{aligned} 
         \pd{u_d^\infty}{y}{2} + \pd{u_d^\infty}{z}{2} &= 0     , \\
         u_d^\infty &= 0                \,\,\,\,\text{if } y=1 \text{ or if } y=-1 \text{ and } |z|\geq\gfz{P} , \\
         \pd{u_d^\infty}{y}{} &= - 1    \,\,\,\,\text{if } y=-1 \text{ and } |z|<\gfz{P} .
    \end{aligned}
\end{equation}
The problem given by \eqref{si:eq:udinf} has been solved in closed form \cite{Philip_ZAMP_1972a,Teo2008-pe}, and highlights that $u_d^\infty(y,z)$ is simply the deviation from the Poiseuille profile in the \emph{infinite-grating} problem.

The function $u_2$ satisfies \eqref{si:eq:gov_eq_u} with the no-slip boundary conditions $u_2 = 0$ at $y=\pm{1}$, and the solution is given  by $u_2=-\left[\partial_xp(x)\right]u_P(y)$. Consequently, the following linear combination of $u_P(y)$ and $u_d^\infty(y,z)$ solves \eqref{si:eq:flow_eqns}:
\begin{equation} \label{si:eq:piecewise_sol}
u(x,y,z) =  
    \begin{cases}
        \left[-\pd{p}{x}{}(x)\right]u_P(y) - \left[\gMa+\pd{p}{x}{}(x)\right]u_d^\infty(y,z) & \text{if } |x| < \gfx/2, \\[11pt]
        \left[-\pd{p}{x}{}(x)\right]u_P(y) & \text{if } \gfx/2 \leq |x| \leq 1/2. \\
    \end{cases}
\end{equation}
To determine the pressure gradient term in \eqref{si:eq:piecewise_sol}, we first pose a piecewise pressure field
\begin{equation} \label{si:eq:piecewise_p}
p(x) =  
    \begin{cases}
        p_1(x) & \text{if } |x| < \gfx/2, \\
        p_2(x) & \text{if } \gfx/2 \leq |x| \leq 1/2. \\
    \end{cases}
\end{equation}
Integrating the continuity equation \eqref{si:eqnondim:cont} across any cross section of the domain shows that the volumetric flow rate $Q = \int_{-P/2}^{P/2}\int_{-1}^{1}u(x,y,z)\,\text{d} y\,\text{d} z$ remains constant in $x$. Further integrating the piecewise solution \eqref{si:eq:piecewise_sol} and invoking \eqref{si:eq:piecewise_p}, we obtain two expressions for the flow rates
\begin{subequations} 
    \begin{align*} 
         Q_1 = \int_{-P/2}^{P/2}\int_{-1}^{1}u_1(x,y,z)\,\text{d}y\,\text{d}z &= \left[-\pd{p_1}{x}{}(x)\right]\dfrac{2P}{3} - \left[\gMa+\pd{p_1}{x}{}(x)\right]\Qdinf, \\
         Q_2 = \int_{-P/2}^{P/2}\int_{-1}^{1}u_2(x,y,z)\,\text{d}y\,\text{d}z &= \left[-\pd{p_2}{x}{}\right]\dfrac{2P}{3},
    \end{align*}
\end{subequations}
where $2P/3$ and $\Qdinf(\gfz,P)$ are the flow rates given by $u_P(y)$ and $u_d^\infty(y,z)$, respectively. Since $Q_2$ must be constant in $x$, the pressure gradient $\partial_x{p_2}$ is necessarily independent of $x$ as well. Equating $Q_1=Q_2$ yields a relationship between the two pressure gradients,
\begin{equation} \label{si:eq:p1}
         \pd{p_1}{x}{}(x) = \left[\dfrac{2P}{2P+3\Qdinf}\right] \pd{p_2}{x}{} - \left[\dfrac{3\Qdinf}{2P+3\Qdinf}\right] \gMa.
\end{equation}

The last condition that must be satisfied by the solution is the fixed pressure drop across the domain given by \eqref{si:eqdim:period_p}. The nondimensional version of this equation, taking $\alpha=1$ and $\beta=0$ in \eqref{si:eqdim:period_p}, leads to $p(x)=p(x+1)+1$. This equation can be made specific to $x=-1/2$ and recast into an integral equation for the gradient
\begin{equation*} 
         \int_{-1/2}^{1/2}\pd{p}{x}{}(x)\,\mathrm{d}x = -1
\end{equation*}
which, after applying the decomposition \eqref{si:eq:piecewise_p}, leads to 
\begin{equation} \label{si:eq:p_jump} 
         \int_{-1/2}^{-\gfx/2}\pd{p_2}{x}{}\,\mathrm{d}x + \int_{-\gfx/2}^{\gfx/2}\pd{p_1}{x}{}(x)\,\mathrm{d}x + \int_{\gfx/2}^{1/2}\pd{p_2}{x}{}\,\mathrm{d}x = -1.
\end{equation}
Substituting \eqref{si:eq:p1} into \eqref{si:eq:p_jump}, we arrive at
\begin{equation} \label{si:eq:p_grads}
     \pd{p_1}{x}{}(x) = - \dfrac{2P+3\Qdinf(1-\gfx)\gMaavg}{2P+3\Qdinf(1-\gfx)} + \dfrac{3\Qdinf}{2P+3\Qdinf}(\gMaavg-\gMa), \qquad\quad
     \pd{p_2}{x}{}    = - \dfrac{2P+3\Qdinf(1-\gfx\gMaavg)}{2P+3\Qdinf(1-\gfx)},
\end{equation}
which, after defining $q_d^\infty = 3Q_d^\infty/(2P)$, can finally be substituted into \eqref{si:eq:piecewise_sol} to produce the closed form solution for the flow field Eq.~1 in the main text. The term $\gMaavg$ in \eqref{si:eq:p_grads} represents the average value of $\gMa$ at the interface, i.e. $\gMaavg = \frac{1}{\gfx} \int_{-\gfx/2}^{\gfx/2}\gMa\,\mathrm{d}x$.

Once the leading-order velocity field \eqref{si:eq:piecewise_sol} is fully determined from known parameters, the relevant quantities characterizing the performance of the SHS can be readily obtained. The \emph{local} centerline slip velocity $\uIc=u(x,y=-1,z=0)$ is
\begin{equation} \label{si:eq:uIc_full} 
         \uIc(x) = 2P\left[\dfrac{(1-\gMaavg)}{2P+3\Qdinf(1-\gfx)}+\dfrac{\gMaavg-\gMa}{2P+3\Qdinf}\right]\uIcinf,
\end{equation}
with $\uIcinf(\gfz,P)=u_d^\infty(y=-1,z=0)$. With the additional assumption of a uniform shear stress $\gMa=\gMaavg$, justified in Section~\textit{Scaling theory for surfactant transport}
, \eqref{si:eq:uIc_full} further simplifies to
\begin{equation} \label{si:eq:uIc_simplified} 
         \uIc = \left[\dfrac{2P\uIcinf}{2P+3\Qdinf(1-\gfx)}\right](1-\gMaavg)\vcentcolon=\uIcc\left(1-\gMaavg\right),
\end{equation}
where we define $\uIcc(\gfx,\gfz,P)$ as the centerline slip velocity for the finite-grating \emph{clean}  case (i.e. $\gMa=\gMaavg=0$). We show in Section \textit{Scaling theory for surfactant transport} 
 that we can model the average shear stress $\gMaavg$ using a scaling analysis of the surfactant transport equations, leading to \equaref{si:eq:model_gMa}. Introducing \eqref{si:eq:model_gMa} into \eqref{si:eq:uIc_simplified}, we arrive at Eq.~3 of the main text. Another common, \emph{global} measure of SHS performance is the increase in flow rate with respect to that of a Poiseuille flow. Our theory predicts
\begin{equation} \label{si:eq:Qd} 
         Q_d = \int_{-P/2}^{P/2}\int_{-1}^{1}\left[u(x,y,z)-u_P(y)\right] \mathrm{d}y\,\mathrm{d}z = \left[\dfrac{2P\gfx\Qdinf}{2P+3\Qdinf(1-\gfx)}\right](1-\gMaavg) \vcentcolon= \Qdc(1-\gMaavg),
\end{equation}
where we again introduce $\Qdc(\gfx,\gfz,P)$ as the increase in flow rate for the finite-grating, clean problem. Perhaps the most common global quantity sought in theoretical SHS studies is the effective slip length $\lame$. Here $\lame$ is defined as the quantity that yields the same increase $Q_d$ in flow rate if the mixed boundary conditions on $y=-1$ are replaced with a uniform Navier-slip condition $u = \lame\partial_y{u}$. Such a flow yields a solution $u_{\lame}(y)=u_P(y)+\lame(1-y)/(2+\lame)$ and thus an increase in flow rate of $2P\lame/(2+\lame)$ which, when equated to $Q_d$, yields an expression for the slip length
\begin{equation} \label{si:eq:lame} 
         \lame = \dfrac{2Q_d}{2P-Q_d} = \dfrac{2\gfx\Qdinf(1-\gMaavg)}{2P+\left[3-\gfx\left(4-\gMaavg\right)\right]\Qdinf}.
\end{equation}
\subsection*{Infinite-grating problem} \label{si:sec:inf_grating}
The calculation of $\Qdinf$ and $\uIcinf$ for any values of $P$ and $\gfz$ require either the numerical solution of a dual trigonometric series \cite{Teo2008-pe} or the solution of nonlinear algebraic equations involving elliptic integrals and elliptic functions \cite{Philip_ZAMP_1972a}. We will not provide such level of detail here and instead refer the reader to \cite{Philip_ZAMP_1972a,Lauga_Stone_JFM_2003,Teo2008-pe}. However, for completeness we provide the asymptotic limits
\begin{subequations}\label{si:eq:asympt_Qdinf} 
    \begin{alignat}{2}
        \Qdinf &\to \dfrac{P^2}{\pi}\ln\left(\sec\left(\dfrac{\pi\gfz}{2}\right)\right) \quad&&\text{for }P\ll{1},\\[10pt]
        \Qdinf &\to 2\gfz{P} \quad&&\text{for }P\gg{1},
    \end{alignat}
\end{subequations}
and
\begin{subequations}\label{si:eq:asympt_uIcinf} 
    \begin{alignat}{2}
        \uIcinf &\to \dfrac{P}{\pi}\mathrm{arccosh}\left(\sec\left(\dfrac{\pi\gfz}{2}\right)\right) \quad&&\text{for }P\ll{1},\\[10pt]
        \uIcinf &\to 2 \quad&&\text{for }P\gg{1}.
    \end{alignat}
\end{subequations}
Since the values of $\Qdinf$ and $\uIcinf$ are monotonically increasing with $P$, the above limit \eqref{si:eq:asympt_uIcinf} reveals that $\uIcinf<2$. \revised{In addition, the limits \eqref{si:eq:asympt_Qdinf}, \eqref{si:eq:asympt_uIcinf} can be combined to obtain useful approximations for all $P$, which are within 12\% of the exact values if $0.5<\phi_z<0.95$, and are of course most accurate at large or small $P$:}
\begin{subequations}\label{si:eq:asympt_approx} 
    \begin{alignat}{2}
        \revised{
        \Qdinf }&\revised{\approx \left\{ \left[ \dfrac{P^2}{\pi}\ln\left(\sec\left(\dfrac{\pi\gfz}{2}\right)\right) \right]^{n_q}  + \left[ 2\gfz{P} \right]^{n_q} \right\}^{1/n_q}
,\quad n_q \approx -1.46}\\[10pt]
      \revised{
      \uIcinf }&\revised{\approx \left\{ \left[ \dfrac{P}{\pi}\mathrm{arccosh}\left(\sec\left(\dfrac{\pi\gfz}{2}\right)\right) \right]^{n_u} + 2^{n_u}\right\}^{{1}/{n_u}} , \quad n_u \approx -1.21.}
    \end{alignat}
\end{subequations}

\section*{Scaling theory for surfactant transport} \label{si:sec:th_surf}

\subsection*{Full problem} \label{si:sec:th_surf_full}

The analysis of the surfactant transport equations is similar to that in \cite{landel20}, which we recapitulate here in order to clarify the differences between the two-dimensional and  three-dimensional cases. The first assumption of our model for \eqref{si:eqnondim:c}-\eqref{si:eqnondim:nonlin} is that the concentration of surfactant is low enough to ensure a \emph{dilute regime}, that is $k\ll{1}$. We expect this assumption to be the case for most situations in which surfactants are not artificially added, for instance, when unwanted contaminants are naturally present in water (as discussed in \cite{landel20}). Additionally, since the interaction parameter $A$ is typically not large in absolute value, with $|A|\lesssim20$ \cite{Chang_Franses_1995}, it is possible to assume that $k|A|\ll{1}$ as well. The nonlinear terms \eqref{si:eqnondim:kin} and \eqref{si:eqnondim:nonlin} in the governing equations can then be linearized, leading to Henry kinetics, that is $\kin=c_I-\Gam+O(k)+O(k|A|)$ and $\nonlin=1+O(k)+O(k|A|)$. Consequently, at leading order in $k$ and $k|A|$, \eqref{si:eqnondim:Gam}--\eqref{si:eqnondim:nonlin} can be simplified, yielding Eq.~2 in the main text.

Applying an integral average to \eqref{si:eqnondim:Gam} 
along the spanwise direction, we obtain
\begin{equation} \label{si:eq:avgz_Gam} 
    \pd{\avgz{u_I\Gam}}{x}{} = \dfrac{\eps}{Pe}\pd{\avgz{\Gam}}{x}{2} + \dfrac{Bi}{\eps}(\avgz{c_I}-\avgz{\Gam}),
\end{equation}
where the spanwise average across the plastron of a function $f(z)$ is defined as $\avgz{f}=\frac{1}{\gfz{P}}\int_{-\gfz{P}/2}^{\gfz{P}/2}f(z)\,\mathrm{d}z$, and where the terms in \eqref{si:eqnondim:Gam} associated with derivatives in $z$ vanish due to the no-slip ($w=0$) and no-flux ($\partial_{z}\Gam=0$) boundary conditions at the edges $z=\pm\gfz{P}/2$ of the plastron. If \eqref{si:eq:avgz_Gam} is further integrated from $x=-\gfx/2$ to $x=\gfx/2$ and boundary conditions $u=0$ and $\partial_x\Gam=0$ are applied at $x=\pm\gfx/2$, we have that 
\begin{equation} \label{si:eq:zero_kin}
    \int_{-\gfx/2}^{\gfx/2}\left(\avgz{c_I}-\avgz{\Gam}\right)\mathrm{d}z = 0, 
\end{equation}
and thus, by virtue of the mean value theorem, an equilibrium condition $\avgz{c_I}=\avgz{\Gam}$ must occur at some coordinate along the interface, which we call $x_0$. Downstream from $x_0$, the flow advection promotes the accumulation of interfacial surfactant, which in turn triggers a net desorption flux and an increase in bulk surfactant with respect to the background level. Upstream from $x_0$, the situation is the opposite, with a deficit of $\Gam$ and $c_I$ with respect to the equilibrium values and a net adsorption flux. Figure~1C,~D depicts this physical scenario with the two distinct regions along the interface.

The second main assumption is to consider the interfacial concentration $\Gam$ as {approximately linear}. In this case, \eqref{si:eq:zero_kin} implies that the equilibrium point must be approximately at the center of the interface (i.e. $x_0\approx{0}$), and thus the bulk concentration at $x_0$ is approximately the background concentration and we have $\avgz{c}(x_0)=\avgz{\Gam}(x_0)\approx{1}$. Consequently, this assumption allows to scale the concentrations at both ends of the interface $x=\pm\gfx/2$ as 
\begin{subequations} 
    \begin{align} \label{si:eq:scales_c_Gam}
         c(x=\pm\gfx/2)    \sim 1 \pm \Delta{c}, \\
         \Gam(x=\pm\gfx/2) \sim 1 \pm \Delta{\Gam},
    \end{align}
\end{subequations}
with $\Delta{c}$ and $\Delta{\Gam}$ the characteristic variation of the concentrations (see Fig.~1C,~D). Additionally, note that an approximately linear $\Gam$ also implies, from \eqref{si:eqnondim:Ma_u}, that the Marangoni shear at the interface is {approximately constant} (i.e.~$\gMa\approx\gMaavg$). This assumption is expected to hold as long as the flow is not in the so-called stagnant cap regime \cite{landel20}, characterized by a strongly nonuniform interfacial concentration. The stagnant cap regime is reached when advection at the interface overcomes both diffusion and kinetic effects \cite{Leal07}, that is, when $\eps{Pe_I}\gg{1}$ and either $Bi/\eps\ll{1}$ or $Da\gg{1}$ \cite{landel20}. Given the typical parameter values in small-scale flows like the ones considered in this study (see Section~\textit{Estimate of surfactant parameters} and Table~\ref{si:tab:nondim_numbers}), we conclude that for long gratings we have $\eps{Pe_I}\lesssim{1}$, which justifies the assumption of an approximately linear $\Gam$. Furthermore, we verify a posteriori that our simulation results show an approximately linear profile of $\Gam$ (see Section \textit{Finite-element simulations}).

Using these two assumptions, it is possible to use scaling arguments on \eqref{si:eqnondim} to obtain an expression for $\gMaavg$ as a function of the nondimensional groups of the problem. We start by scaling the terms in \eqref{si:eqnondim:Ma_u} as $\left.\partial_y{u}\right|_I\sim\gMaavg$ and $\partial_x\Gam\sim\Delta{\Gam}/\gfx$, leading to 
\begin{equation} \label{si:eq:scaling_DeltaGam}
    \Delta\Gam \sim \dfrac{\gfx\gMaavg}{\eps{k}{Ma}}. 
\end{equation}
Next, we evaluate the terms in \eqref{si:eqnondim:kin} at the interface ends $x=\pm\gfx/2$. We take $\left.\partial_{y}c\right|_I\sim[1-(1\pm\Delta{c_I})]/\bl\sim\mp\Delta{c_I}/\bl$ and $(c_I-\Gam)\sim[1\pm{\Delta{c_I}}-(1\pm\Delta\Gam)]\sim\pm(\Delta{c_I}-\Delta\Gam)$, where $\bl=\hat{\delta}/\hh$ is the characteristic boundary layer thickness of the bulk concentration (Fig.~1C). We arrive at
\begin{equation} \label{si:eq:scaling_Deltac}
    \Delta{c_I} \sim \dfrac{\bl{Da}}{(1+\bl{Da})}\dfrac{\gfx\gMaavg}{\eps{k}{Ma}}. 
\end{equation}
\equaref{si:eq:avgz_Gam} is integrated from $x=-\gfx/2$ to $x=x_0$, leading to
\begin{equation} \label{si:eq:avgz_Gamma_integrated}
    \avgz{u_I\Gam}(x_0) = \dfrac{\eps}{Pe_I}\pd{\avgz{\Gam}}{x}{}(x_0) + \dfrac{Bi}{\eps}\int_{-\gfx/2}^{x_0}\left(\avgz{c_I}-\avgz{\Gam}\right)\mathrm{d}x,
\end{equation}
whose terms we scale as $\avgz{u_I\Gam}(x_0)\sim\uIc$, $\partial_x{\avgz{\Gam}}(x_0)\sim\Delta\Gam/\gfx$, and $\int_{-\gfx/2}^{x_0}\left(\avgz{c_I}-\avgz{\Gam}\right)\mathrm{d}x \sim \gfx(\Delta{c_I}-\Delta{\Gam})$. Using \eqref{si:eq:scaling_DeltaGam} and \eqref{si:eq:scaling_Deltac} and introducing $g=\hg/\hh=\gfx/\eps$, we arrive at
\begin{equation*}
    \uIc \sim \dfrac{\gMaavg}{kMa}\left(\dfrac{1}{Pe_I}+\dfrac{Bi\,g^2}{(1+\bl{Da})}\right),
\end{equation*}
which, after introducing empirical coefficients for each term, yields
\begin{equation} \label{si:eq:model_uIc}
    \uIc = \dfrac{\gMaavg}{a_{1}kMa}\left(\dfrac{1}{Pe_I}+a_2\dfrac{Bi\,g^2}{(1+\bl{Da})}\right).
\end{equation}
Making use of the theory for the flow field [\ref{si:eq:uIc_simplified}], we substitute $\uIc=(1-\gMaavg)\uIcc$ into \eqref{si:eq:model_uIc} and obtain the expression for $\gMaavg$ as a function of the parameters of the problem,
\begin{equation} \label{si:eq:model_gMa}
    \gMaavg = \dfrac{a_{1}\,k\,Ma\,\uIcc}{\dfrac{1}{Pe_I}+a_{1}\,k\,Ma\,\uIcc+a_2\dfrac{Bi\,g^2}{(1+\bl{Da})}}.
\end{equation}
Equation~\ref{si:eq:model_gMa} can now be introduced in \eqref{si:eq:uIc_full} to obtain the formula for the slip velocity [3] in the main text. Similarly, combining \eqref{si:eq:model_gMa} with \eqref{si:eq:Qd} and \eqref{si:eq:lame}, expressions for the increase in flow rate and effective slip length can be reached.

The only yet undetermined part of the model is an expression for the boundary layer thickness $\bl$, which we seek through scaling of the conservation law for the bulk surfactant \eqref{si:eqnondim:c}. In situations with $\eps{Pe}\gg{1}$, streamwise advection must balance wall-normal diffusion $u\partial_x{c}\sim\frac{1}{\eps{Pe}}\partial_{yy}c$, which is only possible if $c$ varies over a small length scale $\bl\ll{1}$ \cite{Leal07}. We take $\partial_{x}c\sim\Delta{c_I}/\gfx$, $\partial_{yy}c\sim\Delta{c_I}/\bl^2$ and the velocity inside the boundary layer as $u\sim\uIc+\gMaavg\bl$. In the case of an interface close to immobilization, i.e. $\uIc\sim{0}$ and $\gMaavg\sim{1}$, these scalings indicate that $\delta\sim(Pe/g)^{-1/3}$ when $\eps{Pe}\gg{1}$. In the opposite case of $\eps{Pe}\ll{1}$, \equaref{si:eqnondim:c} is dominated by diffusion, and thus the characteristic length scale of variation of $c$ in the wall-normal direction is the whole half height of the domain, implying $\bl\sim{1}$. We choose 
\begin{equation}\label{si:eq:bl}
    \bl=a_3(1+a_4{Pe}/g)^{-1/3}
\end{equation}
to satisfy these two extremes, with $a_3$ and $a_4$ empirical parameters. It is also possible to obtain a similar expression with an exponent of $-1/2$ instead, by assuming that the boundary layer is essentially shear-free (i.e. $\uIc\sim{1}$ and $\gMaavg\sim{0}$). In practice, the overall value of quantities like $\uIc$ are only weakly dependent on the specific functional form of $\bl$, so we only consider the expression \eqref{si:eq:bl}. Additionally, in the case of interest of long gratings $\eps\ll{1}$ in small-scale flows we typically have ${Pe}/g\lesssim{1}$ (Section \textit{Estimate of surfactant parameters}) and thus the boundary layer thickness is approximately independent of $Pe$ or $g$.

\subsection*{Insoluble surfactant limit}
\label{si:sec:th_surf_insol}

All previous theoretical expressions can also be obtained in the case of an insoluble surfactant, i.e. taking $\kin=0$ in \eqref{si:eqnondim:ads-des} and neglecting \eqref{si:eqnondim:c} and \eqref{si:eqnondim:kin}. In this case, $\hGamz$ is an independent parameter that is not linked to $\hcz$, since the bulk concentration is undefined for an insoluble surfactant. The value of $k$ is now simply $k=\hGamz/\hGamm$, although we assume $k\ll{1}$ still holds and leads to $\nonlin\approx{1}$. Furthermore, since $\eps{Pe_I}\lesssim{1}$ remains valid we can still assume a regime away from the stagnant cap and thus an approximately linear profile for $\Gam$. The same steps taken for the scaling of \eqref{si:eqnondim:Ma_u} and \eqref{si:eq:avgz_Gam} can be followed to arrive at
\begin{subequations}
    \begin{align*}
        \gMaavg &= \dfrac{a_\text{ins}\,Ma_\text{ins}\,\uIcc}{1+a_\text{ins}\,Ma_\text{ins}\,\uIcc},\\
        \uIc &= \dfrac{\uIcc}{1+a_\text{ins}\,Ma_\text{ins}\,\uIcc},\\
       \revised{ \lame} & \revised{= \dfrac{\lame^\text{clean}}{1+a_\text{ins}\,Ma_\text{ins}\,\uIcc\left(1+\frac{\lame^\text{clean}}{2} \right)},}
    \end{align*} 
\end{subequations}
with $Ma_\text{ins} = k\,Ma\,Pe_I = \ns\hR\hT\hGamz\hh/(\visc\hDI)$, and $a_\text{ins}$ another empirical parameter.

\section*{Finite-element simulations} \label{si:sec:fe_sims}

We solve numerically the full governing equations \eqref{si:eqdim:cont}-\eqref{si:eqdim:no_fl_Gz} of the problem in dimensional form, performing a total of 155 simulations. The objectives are to (i) determine the values of the empirical parameters $a_1$, $a_2$, $a_3$ and $a_4$ in our model (Section \textit{Scaling theory for surfactant transport}), (ii) confirm the modeling assumptions, for the interfacial concentration $\Gam$, of an approximately constant profile in the spanwise direction, and of a linear profile in the streamwise direction (Section \textit{Scaling theory for surfactant transport}), and (iii) compare the theory to simulations of realistic microchannels in conditions representative of our experiments (Section \textit{Experimental methods}).

\begin{figure*}[!ht]
    \centering
    \captionsetup{farskip=0pt, nearskip=0pt}
    \subfloat{\includegraphics[]{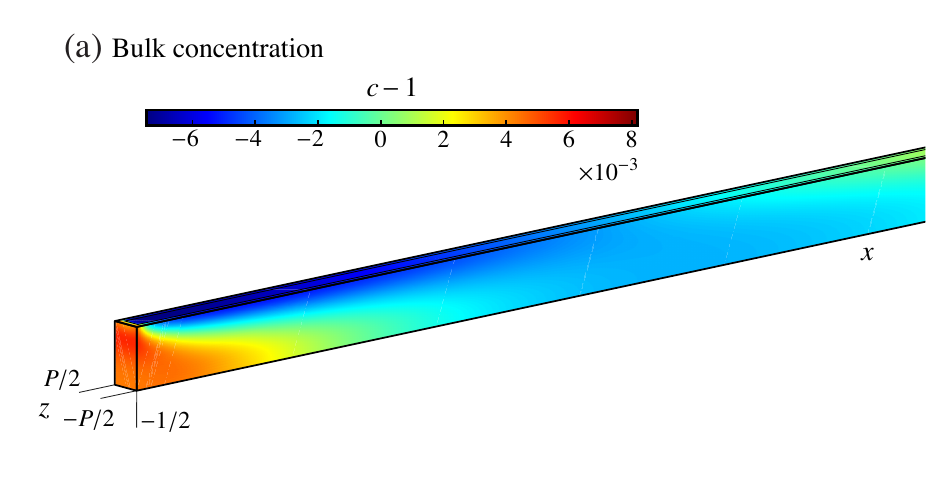}\label{si:fig:sim_bulk}}
    \subfloat{\includegraphics[]{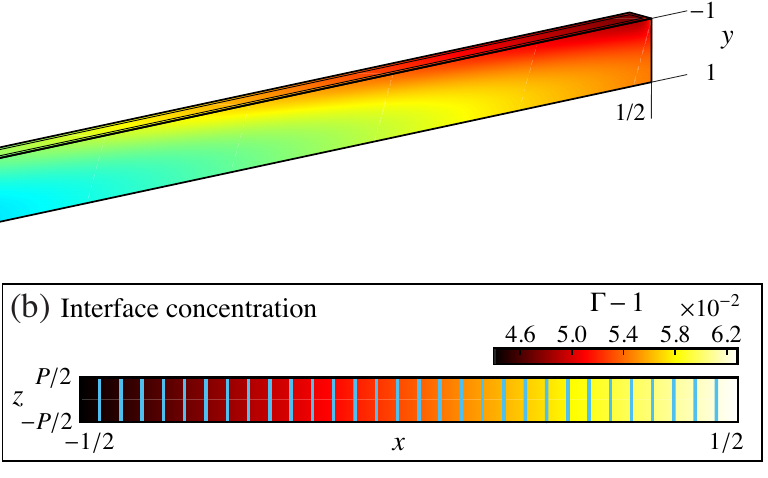}\label{si:fig:sim_intf}}\\[-1pt]
    \subfloat{\includegraphics[]{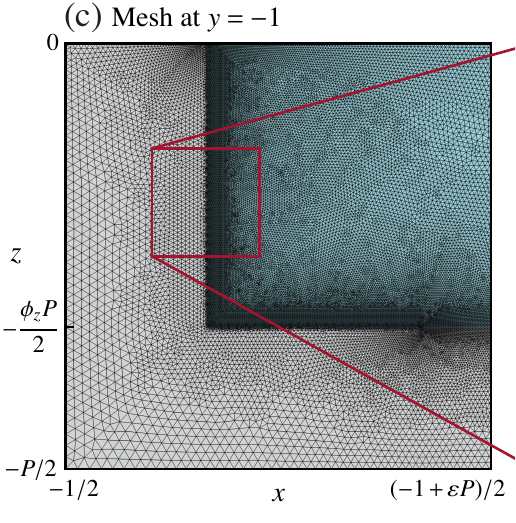}\label{si:fig:mesh}}
    \subfloat{\includegraphics[]{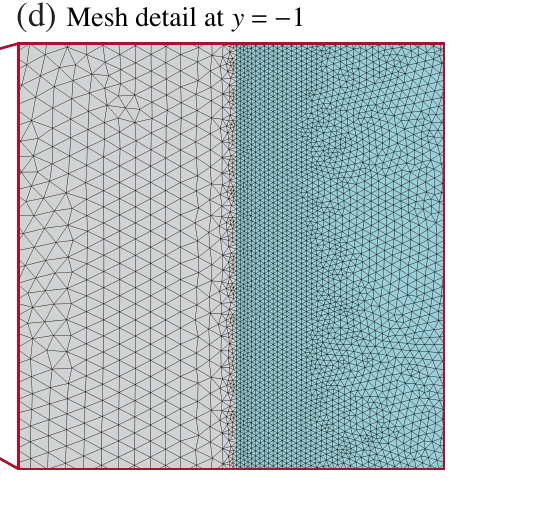}\label{si:fig:mesh_zoom}}
    \subfloat{\includegraphics[]{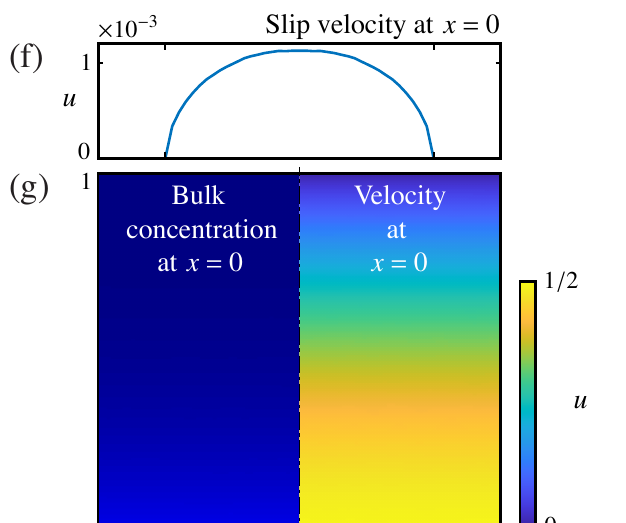}\label{si:fig:slip_vel}}\\[-1pt]
    \subfloat{\includegraphics[]{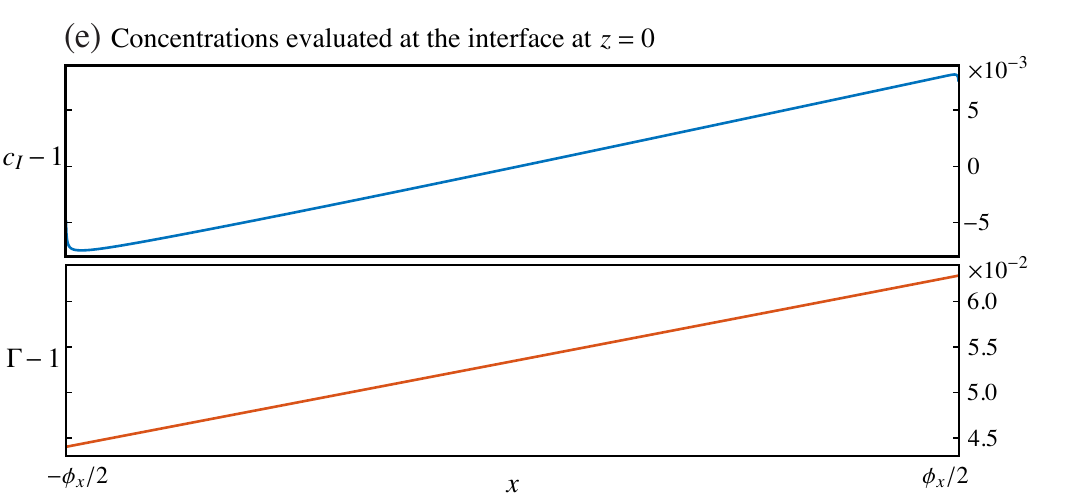}\label{si:fig:concentrations}}
    \subfloat{\includegraphics[]{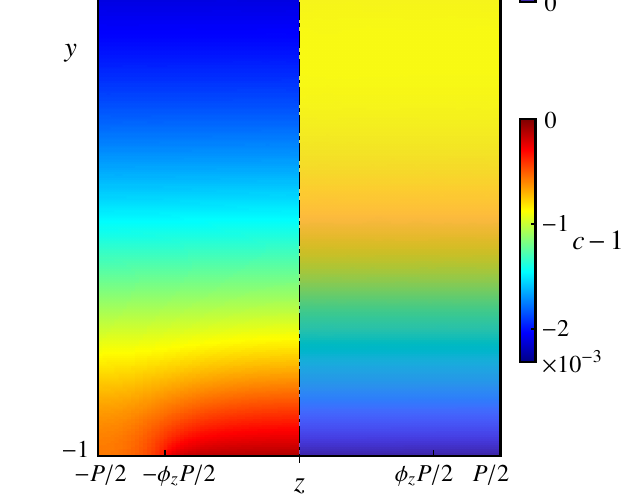}\label{si:fig:cross_section}}
    \caption{Results from the finite-element numerical simulations, obtained with the parameter values that were estimated for the experiments (third column of Table \ref{si:tab:dim_numbers}), using a grating length of $\hg=\SI{3.5}{m\meter}$ due to limitations in computing power. Note that the assumption of a spanwise constant interface concentration is satisfied as shown in (b). In addition, the profiles of $c_I$ and $\Gamma$ remain approximately linear as shown in (e). Despite the nonzero slip velocity shown in (f), the velocity profile is very close to a purely parabolic Poiseuille flow in (g), since at this grating length the interface is nearly immobilized. \label{si:fig:FEsims}}
\end{figure*}

We implemented the three-dimensional simulations using the finite-element software COMSOL Multiphysics 5.5\textsuperscript{\textregistered}. The simulation domain is one half of the SHS unit cell depicted in Fig.~1, with $\hz$ spanning only between $\hz=0$ and $\hz=\hP/2$ due to the spanwise symmetry of the solution. The volume is meshed with tetrahedral elements, concentrating the finest regions around the upstream and downstream edges of the interface $\hx=\pm\gfx\hL/2$ since it is in those areas where the most abrupt variations of the solution occur (see \figref{si:fig:FEsims}). Across all the simulations, the minimum element size (understood as the diameter of a sphere circumscribing the smallest element) is set to $\SI{1.5e-9}{\meter}$. 

The solution of the governing equations is achieved with a combination of the Creeping Flow module for the flow equations \eqref{si:eqdim:cont}-\eqref{si:eqdim:w} and the Dilute Species Transport module for the transport of bulk surfactant \eqref{si:eqdim:c}. The conservation law for the interfacial surfactant \eqref{si:eqdim:Gam} is implemented through a General Form Boundary PDE, using \eqref{si:eqdim:kin} as the source term. The Marangoni boundary conditions \eqref{si:eqdim:Ma_u} and \eqref{si:eqnondim:Ma_w} are enforced through a Weak Contribution constraint, as is the condition that fixes the mean bulk concentration \eqref{si:eqdim:c_zero}.

The system of nonlinear equations is solved through a Newton-type iterative method using the PARDISO direct solver for the linear system at each iteration. We set a relative tolerance of $10^{-5}$ as a convergence criterion for the solution, which is satisfied by all of our simulations. The pressure, bulk concentration and interfacial concentration are discretized using linear elements, and the velocity field uses either quadratic or linear elements, depending on the computational demands of each simulation.

We vary the problem parameters to ensure that each of the distinct terms that are pre-multiplied by an empirical coefficient in \eqref{si:eq:model_gMa} changes over a few orders of magnitude. The ranges of variation of each dimensional quantity in the simulations, as well as of the corresponding nondimensional numbers, is indicated in Tables~\ref{si:tab:dim_numbers} and \ref{si:tab:nondim_numbers}. A small number of simulations were chosen with the same parameters as those estimated in the experiments, in order to achieve a direct comparison (see Fig.~3 in the main text). However, due to constraints in computational power, the value of the grating length $\hg$ was smaller than that of the microfluidic devices.

The parameters $a_1$, $a_2$, $a_3$ and $a_4$ are obtained through least-squares fitting using the MATLAB function \texttt{lsqnonlin}. We define the error as the total sum of squares of the difference between the centerline slip velocities computed in the simulations and those predicted by the theory [3], i.e. $\text{ERR} = \sum(\uIc^\text{theory}-\uIc^\text{sim})^2$. We find $a_1=0.345$, $a_2=0.275$, $a_3=5.581$, and $a_4=3.922$. As illustrated in Fig.~2A, the agreement between simulations and theory is excellent over more than four orders of magnitude in the slip velocity.

\section*{Experimental methods} \label{si:sec:exp_methods}

The experimental setup is centered around the custom-built microfluidic device depicted in Fig.~3A--C. The chips are made by casting PDMS (Sylgard 184) with a 1:10 mass ratio of elastomer-to-curing agent, using a master mold fabricated by two-layer photolithography. The photoresist used for the mold is SU-8 (Microchem SU-8 3050 for the first layer, corresponding to the main channel, and Microchem SU-3025 for the second layer, corresponding to the gratings). \revised{The flat PDMS substrate is determined to be hydrophobic, with static contact angles of water droplets measured to be higher than 90\textdegree. The textured PDMS exhibits a further increase of the static contact angle, with air pockets within the texture that are observable under the microscope, demonstrating that the PDMS textured substrate is in a Cassie-Baxter superhydrophobic state. Measured values of static contact angles are consistent with previous measurements for untreated PDMS \cite{Miranda:22}. }

The chip is bonded to a $\SI{0.1}{m\meter}$-thick glass coverslip (Bellco Glass 1916-25075) through untreated adhesion, and a 40X water objective is used to image the interior of the channels through the coverslip using a confocal microscope (Leica SP8 Resonant Scanning). The device is placed inside a stage top chamber (Okolab H101-K-FRAME) that ensures precise control of temperature, which we set to $\hT=\SI{296}{\kelvin}$. The fluid is initially contained in a glass syringe (Hamilton Gastight), and driven by a syringe pump (KD Legato 111) at a constant flow rate through plastic tubing (Tygon S3) into and out of the microfluidic channel. We use the barrel of a plastic syringe (BD Luer-Lok) as an outlet reservoir open to the room. 
To ensure that the air-water interface in the observed channels remains flat and that plastron curvature effects can be safely neglected, the magnitude of the pressure inside the channel is adjusted by varying the height of the outlet reservoir, which is mounted on a vertical translation stage (Thorlabs VAP10). The maximum deflection of the interface at the centerline, relative to the edges, is estimated to be less than $\pm\SI{1}{\micro \metre}$.

Due to the extreme difficulty of removing all traces of surface-active contaminants, even in controlled experimental conditions \cite{peaudecerf17}, we do not attempt an exhaustive cleaning protocol with that aim. Nevertheless, we follow standard cleaning procedures on all syringes and tubing, ensuring that they are rinsed with $\SI{18}{\mega\ohm\centi\meter}$ DI water with at least twice their volume before they are used. In addition, we follow a specific cleaning protocol for the \uPIV~particles (ThermoFisher FluoSpheres carboxylate 0.5-\SI{}{\micro\meter} diameter), since they typically contain surfactants to prevent particle aggregation \cite{Li20}. We use a centrifuge (Eppendorf 5418) to separate the beads from the buffer solution, which is discarded and replenished with clean $\SI{18}{\mega\ohm\centi\meter}$ DI water, and we repeat the process three times. These cleaning procedures ensure that the traces of surfactants responsible for the non-negligible Marangoni stresses that we observe in the experiments are the result of contamination that would naturally occur in typical small-scale flows through microfluidic devices, and not as a byproduct of the specific experimental methods used in this study.

The \uPIV~analysis is performed using the open-source MATLAB toolbox \texttt{PIVlab}. The acquisition window has an approximate size of $\SI{125}{\micro\meter}\times\SI{125}{\micro\meter}$, which is sufficient to cover the span of two pitches (see Fig.~3A,B,C), which are imaged around the center of the grating in the streamwise direction (i.e. $x=0$). We image the motion of the \uPIV~particles across time intervals between $\SI{20}{\second}$ and $\SI{60}{\second}$, at different distances from the interface, with frame rates between $\SI{20}{fps}$ and $\SI{28}{fps}$.The velocity field is averaged in time and in the streamwise ($x$) direction to obtain the velocity profiles depicted in Fig.~3D. To calculate the value of $\huIc$, we perform a linear least-squares fit, typically using between three and five velocity profiles to obtain an extrapolated slip velocity, from which we extract its value at $z=0$. This linear fit is performed in MATLAB with a script that takes into account the propagation of uncertainties in the distance $\Delta\hat{y}$ from the interface, as well as the uncertainty in $\hat{u}$ inherent to the measurement and the averaging in the $x$ direction.

\section*{Estimate of surfactant parameters} \label{si:sec:estimate}

The main challenge in comparing the experimental measurements of $\huIc$ to the predictions from our model is the absence of information regarding the type and amount of surfactant present in the channels. Some parameters in the problem are known from the experimental conditions, and hence we fix those as $\ns=2$ \cite{prosser01}, $\hR=\SI{8.314}{\joule\per\mole\per\kelvin}$, $\hT=\SI{296}{\kelvin}$ and $\visc=\SI{8.9e-4}{\kilogram\per\meter\per\second}$ \cite{crc:14}. Others can be accurately estimated, since most surfactants have diffusivities (both $\hD$ and $\hDI$) bounded between \SI{5e-10}{} and \SI{5e-9}{\square\meter\per\second}, and have values of $\hGamm$ between \SI{e-6}{} and \SI{5e-5}{\mole\per\square\meter} \cite{Chang_Franses_1995}. We thus use as a reference surfactant the well-studied sodium dodecyl sulfate (SDS), setting $\hD=\hDI=\SI{7e-10}{\square\meter\per\second}$ and $\hGamm=\SI{3.9e-6}{\mole\per\square\meter}$. However, the values of the rate constants $\hka$ and $\hkd$ can vary significantly across surfactants \cite{Chang_Franses_1995}, and they cannot be assumed a priori. In addition, we must estimate the background bulk concentration $\hcz$.

Here we describe a strategy to estimate these quantities. The value of $k = \hka \hcz/\hka$ has an upper bound, since we expect $k\ll{1}$ not only because this is the case when surfactants are not artificially added, but also because values of $k\gtrsim{1}$ would significantly decrease the mean surface tension, leading to a rapid plastron collapse, which is not observed in experiments. We therefore choose the bound $k<k_\text{max}=\SI{e-1}{}$, which ensures that $k$ remains at least one order of magnitude smaller than 1 and that the absolute surface tension decrease $\Delta\hat{\sigma}$ is small compared to the clean surface tension value $\hat{\sigma}_0=\SI{7.2e-2}{\newton\per\meter}$. Indeed, an estimation using an equation of state derived from the Langmuir isotherm \cite{landel20} yields $\Delta\hat{\sigma}/\hat{\sigma}_0=\ns\hR\hT\hGamm\ln(1+k_\text{max})/\hat{\sigma}_0\approx{0.025}$.

To establish a lower bound for $k$, we note that a near-immobilized plastron, as in the experiments in \cite{peaudecerf17}, requires that $\eps{k}{Ma}\Delta\Gamma/\gfx=k{Ma}\Delta\Gamma/g\approx{1}$, from a scaling analysis of \eqref{si:eqnondim:Ma_u}. In conditions far away from the stagnant cap regime, the increase of interfacial surfactant $\Delta\Gamma$ can also be bounded $\Delta\Gamma<1$, giving $k{Ma}/g\gtrsim{1}$. This leads to a lower bound for $k$ given by $k>k_\text{min}=g/Ma$ which, using the values $g=600$ and $Ma\approx\SI{8.3e4}{}$ (estimated using $\hGamm=\SI{3.9e-6}{\mole\per\square\meter}$ and $\hU\approx{3}\huavg=\SI{2.61e-4}{\meter\per\second}$ from \cite{peaudecerf17}, where $\huavg$ is the velocity average across the $yz$-plane), yields
\begin{equation}
    \SI{7.3e-3}{} \lesssim k \lesssim \SI{e-1}{}.
    \label{si:eq:bounds_k}
\end{equation}

In addition, the expression \eqref{si:eq:model_uIc} can be combined with the quantitative results from \cite{peaudecerf17} to estimate the kinetic rate constants $\hka$ and $\hkd$. For long gratings (i.e.~$g\to\infty$), \eqref{si:eq:model_uIc} indicates that the slip velocity converges to the clean-case value $\uIc\to\uIcc$ (see~Fig.~3F). However, for intermediate lengths in which $\uIc$ is still close to $\uIcc$, the dominant balance of terms in \eqref{si:eq:model_uIc} yields the approximation $\uIc\approx{a_2}Bi\,{g^2}/[a_1{k}Ma(1+\bl{Da})]$. Moreover, the estimated order of magnitude of $\hDI$ results in a thick boundary layer $\bl\approx{a_3}$, since $Pe/g=O(10^{-2})\ll{1}$ when the values $\hh=\SI{5e-5}{\meter}$, $\hU\approx{3}\huavg=\SI{2.61e-4}{\meter\per\second}$ and $g=600$ from \cite{peaudecerf17} are used. The value of $\hka\hGamm$ is generally large enough to guarantee $\hka\hGamm>10^{-5}\,\SI{}{\meter\per\second}$ \cite{Chang_Franses_1995}, which suggests that $Da\gg{1}$. This means that $(1+Da\,\bl)\approx{a_3}Da$, yielding
\begin{equation}
    k\dfrac{\hka}{\hkd}\approx\frac{a_2{g^2}}{{a_1}{a_3}Ma}\left(\frac{\hD}{\huIc\hGamm}\right).
    \label{si:eq:estimate}
\end{equation}
We use $\huIc=12.18\pm$\SI{3.48}{\micro \meter \per\second} from \cite{peaudecerf17} in the right-hand side of \eqref{si:eq:estimate}; the other parameters are already known or estimated. Combining this expression with the bounds \eqref{si:eq:bounds_k} for $k$, we obtain
\begin{equation}
    \SI{7.1e1}{\dfrac{m^3}{mol}} \lesssim \dfrac{\hka}{\hkd } \lesssim \SI{1.8e3}{\dfrac{m^3}{mol}}.
    \label{si:eq:bounds_kin}
\end{equation}
As mentioned previously, the Damk\"ohler number is high for most surfactants \cite{Chang_Franses_1995}, so the values of $\hka$ and $\hkd$ have a weak effect individually, and the ratio $\hka/\hkd$ from \eqref{si:eq:bounds_kin} is enough to characterize the surfactant. The grey band in Fig.~3F corresponds to the bounds set by \eqref{si:eq:bounds_k} and \eqref{si:eq:bounds_kin} in the limit of $Da\to\infty$. The edges of the band change only slightly when values of $Da$ as low as 1 are considered. 

The specific choice of $\hcz=\SI{3e-4}{\mole\per\meter\tothe{3}}$, $\hka=\SI{89.5}{\meter\tothe{3}\per\mole\per\second}$ and $\hkd=\SI{0.75}{\per\second}$ (which leads to $k=\SI{3.58e-2}{}$ and $\hka/\hkd=\SI{1.19e2}{\meter\tothe{3}\per\mole}$) yields a good agreement with our experimental results (Fig.~3F).


\section*{Discussion of the mobilization length} \label{si:sec:L_m}
\begin{figure*}[b!]
    \centering
    \includegraphics{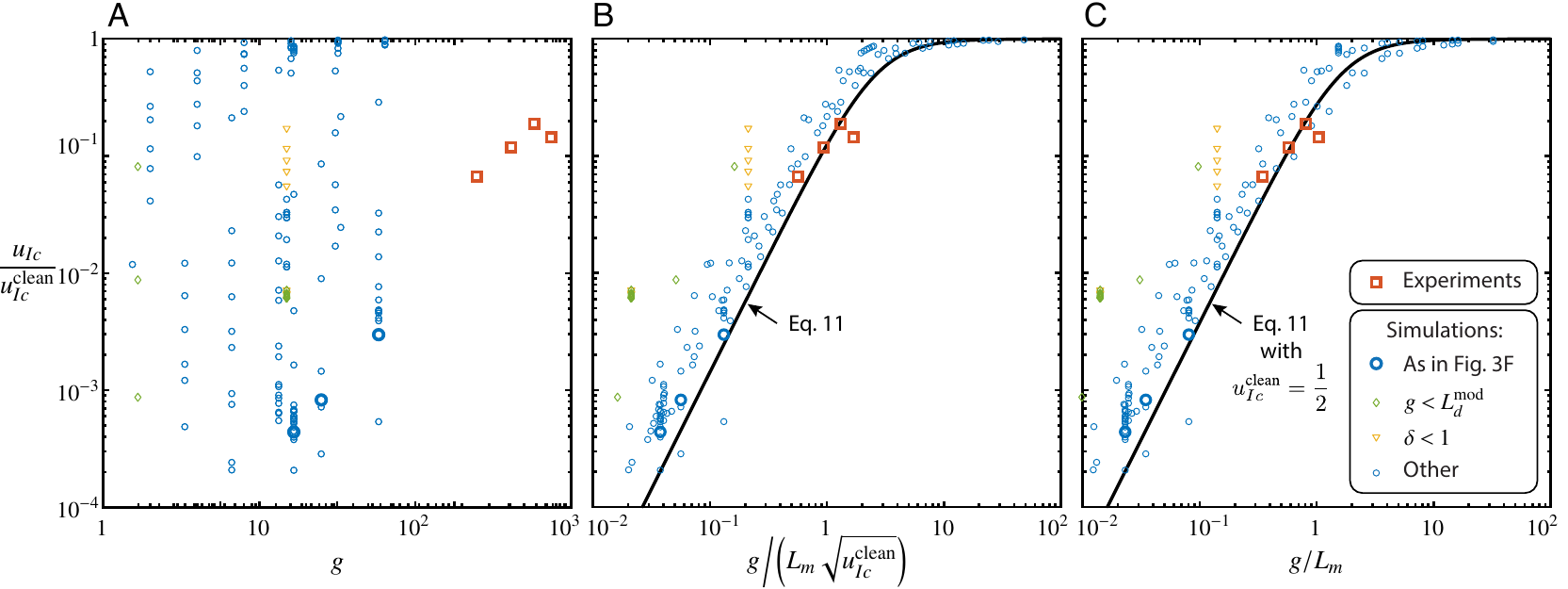}
    \caption{(A) and (B): Collapse of slip velocity data after normalization with the mobilization length (as in Figs.~4B,~C), including all 155 simulations. Yellow triangles are simulations with a thin boundary layer $\delta < 1$, which is not representative of long gratings, whereas the green diamonds denote cases where the grating length is below the modified depletion length $g<\Ldmod$, which are of little practical relevance, as explained in the text. (C) Data collapse when $\uIcc$ is ignored in the normalization of $g$. The curve shows Eq.~11 with a constant value of $\uIcc=1/2$. \label{si:fig:L_m_collapse}}
\end{figure*}

\revised{
Figures \ref{si:fig:L_m_collapse}A,~B display the same data collapse as Figs.~4B,~C, but include the data from all 155 simulations. The yellow triangles indicate simulations with a thin boundary layer $\delta < 1$, which can only occur if $Pe/g\gg{1}$. In practical applications with long gratings $g\gg{1}$, such a regime would require extremely large P\'eclet numbers $Pe\gg{g}\gg{1}$ that are outside the scope of this study. The green diamonds denote cases in which the grating length is below the modified depletion length $g<\Ldmod$, which are of little practical relevance. This is due to the fact that, for typical surfactants, we find $\Lm\gg\Ldmod$ (see Section \textit{Problem parameters}), and thus $\Lm$ is ultimately the lengthscale that $g$ needs to overcome for a SHS grating to display significant slip.
}

\revised{
Figure~\ref{si:fig:L_m_collapse}C shows the same data collapse when the parameter $(\uIcc)^{1/2}$ is omitted from the normalization of $g$. Note that $\uIcc$ depends only on the SHS texture geometry. Since, for SHS capable of drag reduction, one needs $\uIcc$ of order one (see Section \textit{Infinite-grating problem}), the collapse in Fig.~\ref{si:fig:L_m_collapse}C is only marginally worse than in Fig.~\ref{si:fig:L_m_collapse}B. Furthermore, a good approximation to the curve shown in Fig.~\ref{si:fig:L_m_collapse}B is found by using Eq.~11 with $\uIcc=1/2$, as shown in Fig.~\ref{si:fig:L_m_collapse}C.
}

\revised{
For convenience, here we provide a detailed discussion of the assumptions under which the surfactant problem is only a function of $g/L_m$: 
\begin{itemize}
    \item \underline{Dilute surfactant regime}, $k\ll{1}$: This is consistent with environmental traces of surfactant, intrinsic to the microfluidic system, without artificially added surfactant. Furthermore, for values of $k\sim O(1)$ the plastron would risk collapse due the decrease in mean surface tension.   Based on estimates from experimental data~\cite{peaudecerf17}, we take bounds $\SI{e-4}{}\lesssim{k}\lesssim\SI{e-1}{}$ (see Section \textit{Estimate of surfactant parameters}), consistently also with values found in the literature for engineered \cite{carter:20} and natural settings \cite{frossard:19}. 
    \item \underline{Boundary layer extending across the microchannel}, $\delta\sim O(1)$: Long gratings favor a thick boundary layer since it has more available length to develop. Since $\delta \approx a_3(1+a_4(Pe/g))^{-1/3}$,  $Pe/g\lesssim O(1)$ is needed to have $\delta\sim O(1)$. For microchannel flows, the half-height is in the range $\hh\in[10^{-4},10^{-3}]\,\SI{}{\meter}$, and the flow velocity $\hU\in[10^{-5},10^{-3}]\,\SI{}{\meter\per\second}$. Surfactants have diffusivities of order $\hD\sim\SI{e-9}{\meter\squared\per\second}$ or smaller \cite{Chang_Franses_1995}, yielding $Pe = \hat{h}\hat{U}/\hat{D}\in[1,10^{3}]$. Therefore gratings with interface length $g = \hg/\hh \sim 1/\varepsilon \sim 10^2$ the thick boundary layer would be fully developed. For longer gratings, this condition is even more easily satisfied. 
    \item \underline{Kinetics are fast compared to diffusion}, $Da\gtrsim{1}$: For microchannel half-heights and diffusivities as in the points above, and finding the approximate bounds $\hat{\kappa}_a\hat{\Gamma}_m \in [10^{-5}, 10^{-3}]\,\mathrm{m\,s^{-1}}$ in the literature (satisfied for all surfactants except for two in Table 3 of \cite{Chang_Franses_1995}), one has that $Da = \hat{\kappa}_a\hat{\Gamma}_m\hat{h}/\hat{D} \in [1,10^{3}]$. Only for $\hh$ considerably smaller than tens of microns one would have very small values of $Da$.
    \item \underline{Mobilization length larger than the modified depletion length}, $L_m \gtrsim L_d^\mathrm{mod}$: Since $L_d^\mathrm{mod} = \sqrt{Da/(Pe_I Bi)}$, and $L_m = \sqrt{k\,Ma\,Da/Bi}$, to achieve $L_d^\mathrm{mod}  \lesssim L_m$ one needs $k\,Ma\,Pe_I = k\, \ns \hR \hT \hGamm \hh (\visc \hDI)^{-1} \gtrsim 1$.
    We take $k$ in the range described in the points above, $\ns=2$, $\hR = 8.314$\,J/(mol\,K)$^{-1}$, $\hT \sim 300\,$K, $\hGamm\sim 10^{-6}$\,mol\,m$^{-2}$, $\visc = \SI{1e-3}{\kilogram\, (\meter\,\second)^{-1}}$, and $\hDI \sim \SI{1e-9}{\meter^2 \second^{-1}}$. We then find $k\,Ma\,Pe_I \in [5\cdot 10^1, 5\cdot 10^4]$, indicating that $L_d^\mathrm{mod}\lesssim L_m$.
    \item \underline{Uniform shear stress}, $\gMa\approx\gMaavg$: Deviations from this assumption are associated with the `stagnant cap regime', which requires a dominant interface advection term, with negligible kinetics and diffusion \cite{Leal07}. The stagnant cap regime is difficult to achieve for long gratings, since $Bi/\eps\gtrsim{1}$ will be easily satisfied, thereby making interface kinetic fluxes significant in the interfacial transport equation (Eq.~2b in the main text). We assume that the channel half-height is in the range $\hh\in[10^{-4},10^{-3}]\,\SI{}{\meter}$, and that the flow velocity is in the range $\hU\in[10^{-5},10^{-3}]\,\SI{}{\meter\per\second}$. Based on the review of physicochemical properties of surfactants in Ref.~\cite{Chang_Franses_1995} (see their Table 3), we assume $\hkd$ is in the range $\hkd\in[10^{-1},10^{3}]\,\SI{}{\per\second}$, which covers all but one of the surfactants that they describe.
    With these bounds, we find $Bi=\hkd\hh/\hU\in[10^{-2},10^{5}]$. Even with this wide range of parameters, the Biot number remains high enough such that $Bi/\eps$ will be at least of order one for gratings that are practical for drag reduction, where $\eps = \hh/\hL \approx \hh/\hg\sim{10^{-2}}$ or smaller. 
The uniform shear stress assumption is also supported by our simulations, which span a wide range of parameters (see Table~\ref{si:tab:nondim_numbers}). 
\end{itemize}
}

\FloatBarrier

\dataset{Temprano-Coleto_etal_Simulation_Data.xlsx}{Data for finite-element simulations of the full Eqs.~\ref{si:eqdim:cont}--\ref{si:eqdim:no_fl_Gz}.}

\bibliography{SHS3D_biblio-si}